\documentclass[letterpaper,12pt]{JHEP3}
\usepackage{amsmath,amssymb}
\usepackage{latexsym}
\usepackage{epsfig}
\usepackage{cite}
\usepackage[english]{babel}

\newcommand{\X}{X\!\!\cdot\!g}
\newcommand{\Xb}{{\bar X}\!\!\cdot\!g}
\newcommand{\Xp}{\dot{X}\!\!\cdot\!g}
\newcommand{\Xbp}{{\dot{\bar X}}\!\!\cdot\!g}

\newcommand{\eq}{\begin{equation}}
\newcommand{\feq}{\end{equation}}
\newcommand{\eqn}{\begin{eqnarray}}
\newcommand{\feqn}{\end{eqnarray}}
\newcommand{\arr}{\begin{eqnarray*}}
\newcommand{\farr}{\end{eqnarray*}}

\font\mybb=msbm10 at 12pt
\def\bb#1{\hbox{\mybb#1}}

\def\bR {\bb{R}}

\def\bC {\bb{C}}

\title{The timelike half-supersymmetric backgrounds of ${\cal N}=2$, $D=4$
supergravity with Fayet-Iliopoulos gauging}

\author{Dietmar Klemm and Emanuele Zorzan \\
Dipartimento di Fisica dell'Universit\`a di Milano, \\
\hspace*{0.15cm} Via Celoria 16, I-20133 Milano and \\
INFN, Sezione di Milano, Via Celoria 16, I-20133 Milano. \\
}
\preprint{IFUM-953-FT}

\abstract{Subject to some relatively mild assumptions, we derive the complete
form of all timelike half-supersymmetric solutions to ${\cal N}=2$, $D=4$ gauged
supergravity coupled to an arbitrary number of abelian vector multiplets.
This is done using spinorial geometry techniques.
Explicit examples are given for a simple prepotential. Among the solutions,
there are near-horizon geometries of extremal rotating BPS black holes still to be
discovered, with a nontrivial dependence of the scalar fields on one of the horizon
coordinates.
}


\begin{document}

\section{Introduction}

BPS solutions to supergravity theories have played, and continue to play,
an important role in string theory developments. Supersymmetric black holes
represent perhaps one of the most notable examples of this: In presence of a
sufficient amount of supersymmetry, non-renormalization
theorems allow to extrapolate an entropy computation at weak string coupling
(when the system is generically described by a configuration of strings and branes)
to the strong-coupling regime, where a description in terms of a black hole is
valid \cite{Strominger:1996sh}. These entropy calculations have been essential for
our current understanding of black hole microstates.
It is therefore important to dispose of a systematic classification of BPS
solutions, that allows to construct such backgrounds without the necessity to
guess suitable ansaetze. Of particular interest in this context are gauged
supergravities, which are related to supersymmetric field theories by
the AdS/CFT correspondence. While we know by now a broad landscape of BPS
solutions to ungauged supergravities, including many different types of
black holes and black rings \cite{Elvang:2004rt}, only a few of their analogues
in gauged supergravity have been constructed\footnote{Note that some
of these analogues might not exist \cite{Kunduri:2006uh}.}. For instance,
in four dimensions, there should exist rotating black holes in gauged ${\cal N}=8$
supergravity (that admits a truncation to ${\cal N}=2$ gauged supergravity coupled
to three abelian vector multiplets \cite{Cvetic:1999xp}) with four independent
electromagnetic charges. Until now, the only known solutions of this type are the
Kerr-Newman AdS black holes, which correspond to setting the four charges equal,
and the black holes in SO$(4)$ gauged ${\cal N}=4$ supergravity with two pairwise
equal charges \cite{Chong:2004na}.

In this paper, we consider the theory of ${\cal N}=2$, $D=4$ gauged
supergravity coupled to an arbitrary number of abelian vector multiplets,
but with no hypermultiplets (so-called Fayet-Iliopoulos gauging). The
constraints obeyed by backgrounds admitting at least one timelike Killing
spinor were given in \cite{Cacciatori:2008ek}, generalizing the results
for minimal gauged supergravity \cite{Caldarelli:2003pb}.
Although the equations determining the BPS geometries are rather involved,
some explicit solutions of them describing static black holes with nontrivial
scalars turned on have been obtained in \cite{Cacciatori:2009iz}. These black holes
provide a new ground to test the AdS/CFT correspondence: In principle
it should be possible to compute their microscopic entropy using the
recently discovered Chern-Simons-matter theories \cite{Aharony:2008ug}, and to
compare it then with the macroscopic Bekenstein-Hawking result.

Here we go one step further with respect to \cite{Cacciatori:2008ek} and
impose the existence of at least two Killing spinors, so we want to determine
the most general half-supersymmetric configurations\footnote{In five
dimensions, this was done in \cite{Gutowski:2007ai} and \cite{Grover:2008ih}
for the timelike and null cases respectively. Maximally supersymmetric
solutions to four-dimensional ${\cal N}=2$ gauged supergravity were
classified in \cite{Hristov:2009uj}.}. There are several
reasons motivating this:

First of all, it is of special interest to address cases of the AdS$_4$/CFT$_3$
correspondence with less than maximal supersymmetry. For instance, supergravity
vacua with lower supersymmetry may have an interpretation on the CFT side as
vacua with non-zero expectation values of certain operators (spontaneous symmetry
breaking), or as deformations of the CFT (explicit symmetry breaking).

The second point is the attractor mechanism \cite{Ferrara:1995ih,Strominger:1996kf,
Ferrara:1996dd,Ferrara:1996um,Ferrara:1997tw}. While
the BPS attractor flow has been studied extensively for asymptotically
flat black holes, the AdS case was considered only
recently \cite{Cacciatori:2009iz}\footnote{For an analysis of the attractor
mechanism in ${\cal N}=2$, $D=4$ supergravity with SU$(2)$ gauging
cf.~\cite{Huebscher:2007hj}.}.
In order to explore the BPS attractor flow in AdS, one needs the
near-horizon geometry of (possibly rotating) AdS black holes with scalar
fields turned on. In the asymptotically flat case, such near-horizon
geometries are typically fully supersymmetric, whereas, as we shall
see below, in AdS they generically break one half of the supersymmetries.

Furthermore, in gauged supergravity, interesting mathematical structures
appear in the base manifolds of reduced holonomy, over which supersymmetric
spacetimes are fibered. For instance, one can have U$(1)$ holonomy with
torsion \cite{Cacciatori:2008ek} (the torsion coming from the gauging),
Einstein-Weyl spaces \cite{Grover:2009ms} or hyper-K\"ahler torsion
manifolds \cite{Grover:2008jr}, and one might ask how these structures are
modified if one imposes the existence of more than one Killing spinor.

Finally, in minimal ${\cal N}=2$, $D=4$ gauged supergravity, the equations
determining the BPS solutions reduce, under some assumptions, to the equations
of motion following from the gravitational Chern-Simons action \cite{Cacciatori:2004rt}.
While the deeper reason for this remains obscure, it indicates that the
full set of equations actually might be integrable, i.e., it should be possible to
construct a Lax pair for them. Requiring additional supersymmetries can help
to better understand the integrability structure of this system.

The remainder of this paper is organized as follows:
In section \ref{FIsugra}, we briefly review the theory of ${\cal N}=2$,
$D=4$ supergravity with Fayet-Iliopoulos gauging. After that, in \ref{1/2susy},
we impose the existence of a second Killing spinor, obtain the linear system
into which the Killing spinor equations turn, and derive the time-dependence of
this second covariantly constant spinor. Subsequently, the linear system is
solved under some relatively mild assumptions, and the spacetime geometry,
the fluxes as well as a scalar flow equation are obtained.
The reader who is interested only in the final results can skip the technical
details and immediately jump to the summaries in sections \eqref{summary},
\eqref{X-Xb=0}, \eqref{X-Xbneq0} and \eqref{2=12=0}.

\section{${\cal N}=2$, $D=4$ supergravity with Fayet-Iliopoulos gauging}
\label{FIsugra}

We consider ${\cal N}=2$, $D=4$ gauged supergravity coupled to $n_V$ abelian
vector multiplets \cite{Andrianopoli:1996cm}\footnote{Throughout this paper,
we use the notations and conventions of \cite{Vambroes}.}.
Apart from the vierbein $e^a_{\mu}$, the bosonic field content includes the
vectors $A^I_{\mu}$ enumerated by $I=0,\ldots,n_V$, and the complex scalars
$z^{\alpha}$ where $\alpha=1,\ldots,n_V$. These scalars parametrize
a special K\"ahler manifold, i.~e.~, an $n_V$-dimensional
Hodge-K\"ahler manifold that is the base of a symplectic bundle, with the
covariantly holomorphic sections
\begin{equation}
{\cal V} = \left(\begin{array}{c} X^I \\ F_I\end{array}\right)\,, \qquad
{\cal D}_{\bar\alpha}{\cal V} = \partial_{\bar\alpha}{\cal V}-\frac 12
(\partial_{\bar\alpha}{\cal K}){\cal V}=0\,, \label{sympl-vec}
\end{equation}
where ${\cal K}$ is the K\"ahler potential and ${\cal D}$ denotes the
K\"ahler-covariant derivative. ${\cal V}$ obeys the symplectic constraint
\begin{equation}
\langle {\cal V}\,,\bar{\cal V}\rangle = X^I\bar F_I-F_I\bar X^I=i\,.
\end{equation}
To solve this condition, one defines
\begin{equation}
{\cal V}=e^{{\cal K}(z,\bar z)/2}v(z)\,,
\end{equation}
where $v(z)$ is a holomorphic symplectic vector,
\begin{equation}
v(z) = \left(\begin{array}{c} Z^I(z) \\ \frac{\partial}{\partial Z^I}F(Z)
\end{array}\right)\,.
\end{equation}
F is a homogeneous function of degree two, called the prepotential,
whose existence is assumed to obtain the last expression.
The K\"ahler potential is then
\begin{equation}
e^{-{\cal K}(z,\bar z)} = -i\langle v\,,\bar v\rangle\,.
\end{equation}
The matrix ${\cal N}_{IJ}$ determining the coupling between the scalars
$z^{\alpha}$ and the vectors $A^I_{\mu}$ is defined by the relations
\begin{equation}\label{defN}
F_I = {\cal N}_{IJ}X^J\,, \qquad {\cal D}_{\bar\alpha}\bar F_I = {\cal N}_{IJ}
{\cal D}_{\bar\alpha}\bar X^J\,.
\end{equation}
The bosonic action reads
\begin{eqnarray}
e^{-1}{\cal L}_{\text{bos}} &=& \frac 1{16\pi G}R + \frac 14(\text{Im}\,
{\cal N})_{IJ}F^I_{\mu\nu}F^{J\mu\nu} - \frac 18(\text{Re}\,{\cal N})_{IJ}\,e^{-1}
\epsilon^{\mu\nu\rho\sigma}F^I_{\mu\nu}F^J_{\rho\sigma} \nonumber \\
&& -g_{\alpha\bar\beta}\partial_{\mu}z^{\alpha}\partial^{\mu}\bar z^{\bar\beta}
- V\,, \label{action}
\end{eqnarray}
with the scalar potential
\eq
V = -2g^2\xi_I\xi_J[(\text{Im}\,{\cal N})^{-1|IJ}+8\bar X^IX^J]\,,
\label{scal-pot}
\feq
that results from U$(1)$ Fayet-Iliopoulos gauging. Here, $g$ denotes the
gauge coupling and the $\xi_I$ are constants. In what follows, we define
$g_I=g\xi_I$.

The supersymmetry transformations of the gravitini $\psi^i_{\mu}$ ($i=1,2$)
and gaugini $\lambda^{\alpha}_i$ are\footnote{They result from the expressions
given in \cite{Vambroes} by taking ${\vec P}_I=\vec e\,\xi_I$ for the moment
maps (FI gauging), where $\vec e$ denotes a unit vector that can be
chosen to point in the 3-direction without loss of generality. The antiselfdual
parts $F^{-I}$ of the fluxes as well as the $\sigma$-matrices and the
K\"ahler-covariant derivatives $\cal D$ are also given in \cite{Vambroes}.}
\eq
\delta\psi^i_{\mu} = D_{\mu}(\omega)\epsilon^i + ig_IX^I\gamma_{\mu}
{\sigma_3}^{ij}\epsilon_j + \frac14\gamma_{ab}F^{-Iab}\epsilon^{ij}\gamma_{\mu}
\epsilon_j(\text{Im}\,{\cal N})_{IJ}X^J\,, \label{delta-psi}
\feq
\eq
\delta\lambda^{\alpha}_i = -\frac12 g^{\alpha\bar\beta}{\cal D}_{\bar\beta}\bar X^I(\text{Im}\,{\cal N})_{IJ}
F^{-J}_{\mu\nu}\gamma^{\mu\nu}\epsilon_{ij}\epsilon^j + \gamma^{\mu}\partial_{\mu}z^{\alpha}
\epsilon_i - 2ig_I{\sigma_3}_{ij} g^{\alpha\bar\beta}{\cal D}_{\bar\beta}\bar X^I\epsilon^j\,,
\label{delta-lambda}
\feq
where
\eq
D_{\mu}(\omega)\epsilon^i = (\partial_{\mu} + \frac14\omega^{ab}_{\mu}\gamma_{ab})\epsilon^i
+ \frac i2A_{\mu}\epsilon^i + ig_IA^I_{\mu}{\sigma_{3j}}^i\epsilon^j\,.
\feq
Here, $A_{\mu}$ is the gauge field of the K\"ahler U$(1)$,
\eq
A_{\mu} = -\frac i2(\partial_{\alpha}{\cal K}\partial_{\mu}z^{\alpha} -
         \partial_{\bar\alpha}{\cal K}\partial_{\mu}{\bar z}^{\bar\alpha})\,. \label{KaehlerU(1)}
\feq

The most general timelike supersymmetric background of the theory described
above was constructed in \cite{Cacciatori:2008ek}, and is given by
\eq
ds^2 = -4|b|^2(dt+\sigma)^2 + |b|^{-2}(dz^2+e^{2\Phi}dwd\bar w)\ ,
\feq
where the complex function $b(z,w,\bar w)$, the real function $\Phi(z,w,\bar w)$
and the one-form $\sigma=\sigma_wdw+\sigma_{\bar w}d\bar w$, together with the
symplectic section \eqref{sympl-vec}\footnote{Note that also $\sigma$ and
$\cal V$ are independent of $t$.} are determined by the equations
\eq
\partial_z\Phi = 2ig_I\left(\frac{{\bar X}^I}b-\frac{X^I}{\bar b}\right)\ ,
\label{dzPhi}
\feq
\begin{eqnarray}
&&\qquad 4\partial\bar\partial\left(\frac{X^I}{\bar b}-\frac{\bar X^I}b\right) + \partial_z\left[e^{2\Phi}\partial_z
\left(\frac{X^I}{\bar b}-\frac{\bar X^I}b\right)\right]  \label{bianchi} \\
&&-2ig_J\partial_z\left\{e^{2\Phi}\left[|b|^{-2}(\text{Im}\,{\cal N})^{-1|IJ}
+ 2\left(\frac{X^I}{\bar b}+\frac{\bar X^I}b\right)\left(\frac{X^J}{\bar b}+\frac{\bar X^J}b\right)\right]\right\}= 0\,,
\nonumber
\end{eqnarray}
\begin{eqnarray}
&&\qquad 4\partial\bar\partial\left(\frac{F_I}{\bar b}-\frac{\bar F_I}b\right) + \partial_z\left[e^{2\Phi}\partial_z
\left(\frac{F_I}{\bar b}-\frac{\bar F_I}b\right)\right] \nonumber \\
&&-2ig_J\partial_z\left\{e^{2\Phi}\left[|b|^{-2}\text{Re}\,{\cal N}_{IL}(\text{Im}\,{\cal N})^{-1|JL}
+ 2\left(\frac{F_I}{\bar b}+\frac{\bar F_I}b\right)\left(\frac{X^J}{\bar b}+\frac{\bar X^J}b\right)\right]\right\}
\nonumber \\
&&-8ig_I e^{2\Phi}\left[\langle {\cal I}\,,\partial_z {\cal I}\rangle-\frac{g_J}{|b|^2}\left(\frac{X^J}{\bar b}
+\frac{\bar X^J}b\right)\right] = 0\,, \label{maxwell}
\end{eqnarray}
\begin{equation}
2\partial\bar\partial\Phi=e^{2\Phi}\left[ig_I\partial_z\left(\frac{X^I}{\bar b}-\frac{\bar X^I}b\right)
+\frac2{|b|^2}g_Ig_J(\text{Im}\,{\cal N})^{-1|IJ}+4\left(\frac{g_I X^I}{\bar b}+\frac{g_I \bar X^I}b
\right)^2\right]\,, \label{Delta-Phi}
\end{equation}
\begin{equation}
d\sigma + 2\,\star^{(3)}\!\langle{\cal I}\,,d{\cal I}\rangle - \frac i{|b|^2}g_I\left(\frac{\bar X^I}b
+\frac{X^I}{\bar b}\right)e^{2\Phi}dw\wedge d\bar w=0\,. \label{dsigma}
\end{equation}
Here $\star^{(3)}$ is the Hodge star on the three-dimensional base with metric\footnote{Whereas
in the ungauged case, this base space is flat and thus has trivial holonomy, here we have U(1)
holonomy with torsion \cite{Cacciatori:2008ek}.}
\eq
ds_3^2 = dz^2+e^{2\Phi}dwd\bar w\ ,
\feq
and we defined $\partial=\partial_w$, $\bar\partial=\partial_{\bar w}$, as well as
\begin{equation}
{\cal I} = \text{Im}\left({\cal V}/\bar b\right)\ .
\end{equation}
Given $b$, $\Phi$, $\sigma$ and $\cal V$, the fluxes read
\begin{eqnarray}
F^I&=&2(dt+\sigma)\wedge d\left[bX^I+\bar b\bar X^I\right]+|b|^{-2}dz\wedge d\bar w
\left[\bar X^I(\bar\partial\bar b+iA_{\bar w}\bar b)+({\cal D}_{\alpha}X^I)b\bar\partial z^{\alpha}-
\right. \nonumber \\
&&\left. X^I(\bar\partial b-iA_{\bar w}b)-({\cal D}_{\bar\alpha}\bar X^I)\bar b\bar\partial\bar z^{\bar\alpha}
\right]-|b|^{-2}dz\wedge dw\left[\bar X^I(\partial\bar b+iA_w\bar b)+\right. \nonumber \\
&&\left.({\cal D}_{\alpha}X^I)b\partial z^{\alpha}-X^I(\partial b-iA_w b)-({\cal D}_{\bar\alpha}\bar X^I)
\bar b\partial\bar z^{\bar\alpha}\right]- \nonumber \\
&&\frac 12|b|^{-2}e^{2\Phi}dw\wedge d\bar w\left[\bar X^I(\partial_z\bar b+iA_z\bar b)+({\cal D}_{\alpha}
X^I)b\partial_z z^{\alpha}-X^I(\partial_z b-iA_z b)- \right.\nonumber \\
&&\left.({\cal D}_{\bar\alpha}\bar X^I)\bar b\partial_z\bar z^{\bar\alpha}-2ig_J
(\text{Im}\,{\cal N})^{-1|IJ}\right]\,. \label{fluxes}
\end{eqnarray}

If the constraints \eqref{dzPhi}-\eqref{dsigma} are satisfied, the solution admits the Killing spinor
$(\epsilon^1,\epsilon_2)=(1,be_2)$ (cf.~appendix \ref{spinors} for a summary of the essential
information needed to realize spinors in terms of forms).

Before we continue, a short comment on K\"ahler-covariance is in order. Under a K\"ahler
transformation
\eq
{\cal K}\mapsto {\cal K}+f(z^{\alpha})+\bar f(\bar z^{\bar\alpha})\ ,
\feq
the Killing spinors transform as
\eq
\epsilon^i\mapsto e^{(\bar f-f)/4}\epsilon^i\ , \qquad \epsilon_i\mapsto e^{-(\bar f-f)/4}\epsilon_i\ .
\feq
On the other hand, under a U$(1)$ gauge transformation
\eq
A^I_{\mu}\mapsto A^I_{\mu}+\partial_{\mu}\chi^I\ ,
\feq
we have
\eq
\epsilon^1\mapsto e^{-ig_I\chi^I}\epsilon^1\ , \qquad \epsilon_2\mapsto e^{-ig_I\chi^I}\epsilon_2\ .
\feq
Under a combined K\"ahler/U$(1)$ transformation with $ig_I\chi^I=(\bar f-f)/4$, the Killing spinor
representative $(\epsilon^1,\epsilon_2)=(1,be_2)$ is forminvariant; it goes over into $(1,b'e_2)$, with
$b'=e^{-(\bar f-f)/2}b$. One easily checks that the eqns.~\eqref{dzPhi}-\eqref{dsigma} are covariant
under K\"ahler transformations if $b$ is replaced by $b'$. In what follows we sometimes use the
K\"ahler-covariant derivatives of $b$ defined by
\eq
D_{\mu}b = (\partial_{\mu}-iA_{\mu})b\ , \qquad D_{\mu}\bar b = (\partial_{\mu}+iA_{\mu})\bar b\ ,
\feq
as well as $D\equiv  D_w$, $\bar D\equiv D_{\bar w}$. These satisfy
$D'_{\mu}b'=e^{-(\bar f-f)/2}D_{\mu}b$.

\section{Half-supersymmetric backgrounds}
\label{1/2susy}

Let us now investigate the additional conditions satisfied by
half-supersymmetric vacua in the timelike class.
As the stability subgroup of the first
Killing spinor was already used in \cite{Cacciatori:2008ek}
to obtain the eqns.~\eqref{dzPhi}-\eqref{dsigma}, the second one cannot
be simplified anymore, and is thus of the general form
\begin{equation}
\epsilon^1=\alpha1+\beta e_{12}\ ,\qquad \epsilon^2=\gamma1+\delta e_{12}\ ,\qquad
\epsilon_1=\bar\alpha e_1-\bar\beta e_2\ ,\qquad \epsilon_2=\bar\gamma e_1-\bar
\delta e_2\ , \label{2nd-spinor}
\end{equation}
where $\alpha,\beta,\gamma,\delta$ are complex-valued functions.

The conditions coming from an additional Killing spinor are easily obtained by
plugging \eqref{2nd-spinor} into \eqref{delta-psi} and \eqref{delta-lambda} (with
$\delta\psi^i_{\mu}=\delta\lambda^{\alpha}_i=0$), and taking into account the constraints
on the bosonic fields implied by the first Killing spinor $(\epsilon^1,\epsilon_2)=(1,be_2)$,
given in \cite{Cacciatori:2008ek}. This will be done in the following subsection.

\subsection{The linear system}

From the vanishing of the gaugini supersymmetry transformations
(\ref{delta-lambda}) we get
\begin{eqnarray}
\label{htg1}(\bar\beta-b\gamma)\partial_z z^\alpha+2
e^{-\Phi}\sqrt{\frac b{\bar b}}(\bar b\bar\alpha+\delta)\partial z^\alpha
&=&4ig^{\alpha\bar\beta}\mathcal{D}_{\bar\beta}\bar X^Ig_I\gamma\ ,\\
\label{htg2}(\bar b\bar\alpha+\delta)
\partial_zz^\alpha-2e^{-\Phi}\sqrt{\frac{\bar b}b}(\bar\beta-b\gamma)
\bar\partial z^\alpha&=&0\ ,\\
\label{htg3}(b\alpha+\bar\delta)\partial_z
z^\alpha-2e^{-\Phi}\sqrt{\frac b{\bar b}}(\beta-\bar b\bar\gamma)
\partial z^\alpha&=&0\ ,\\
\label{htg4}(\beta-\bar b\bar\gamma)
\partial_zz^\alpha+2e^{-\Phi}\sqrt{\frac{\bar b}b}(b\alpha+\bar\delta)
\bar\partial z^\alpha&=&-\frac{4i}b g^{\alpha\bar\beta}\mathcal{D}_{\bar\beta}
\bar X^Ig_I\beta\ ,
\end{eqnarray}
while the gravitini variations (\ref{delta-psi}) yield
\begin{eqnarray}
\partial_t\alpha&=&-i\bar b\Omega_z(b\alpha+\bar\delta)
+2ie^{-\Phi}|b|\Omega_w(\beta-\bar b\bar\gamma)\ ,\nonumber\\
\partial_t\beta&=&2ie^{-\Phi}\bar b|b|\Omega_{\bar w}(b\alpha+\bar\delta)
+ib\bar b\Omega_z(\beta-\bar b\bar\gamma)
+4i(b\X+\bar b\Xb)\beta-4ib\bar b\X\bar\gamma\ ,\nonumber\\
\partial_t\gamma&=&2i|b|e^{-\Phi}\Omega_w(\bar b\bar\alpha+\delta)+i\bar b
\Omega_z(\bar\beta-b\gamma)+4i\X\bar\beta-4i(b\X+\bar b\Xb)\gamma\ ,\nonumber\\
\label{htgr1}\partial_t\delta&=&ib\bar b\Omega_z(\bar b\bar\alpha+\delta)-
2ie^{-\Phi}\bar b|b|\Omega_{\bar w}(\bar\beta-b\gamma)\ ,
\end{eqnarray}
\begin{align}
\partial_z\alpha&=-\frac{i\Omega_z}{2b}(b\alpha+\bar\delta)
-\frac{ie^{-\Phi}}{|b|}\Omega_w(\beta-\bar b\bar\gamma)\ ,\nonumber \displaybreak[0] \\
\partial_z\beta&=i\sqrt{\frac{\bar b}b}e^{-\Phi}\Omega_{\bar w}(b\alpha+\bar
\delta)-\frac i2\Omega_z(\beta-\bar b\bar\gamma)
+\beta\partial_z\ln|b|+2i\X\bar\gamma\ ,\nonumber \displaybreak[0] \\
\partial_z\gamma&=-\frac{ie^{-\Phi}}{|b|}\Omega_w(\bar b\bar\alpha+\delta)+
\frac i{2b}\Omega_z(\bar\beta-b\gamma)+\frac{2i\X}{b\bar b}\bar\beta-
\frac{\gamma}2\partial_z\ln\frac b{\bar b}\ ,\nonumber \displaybreak[0] \\
\label{htgr2}\partial_z\delta&=-ie^{-\Phi}\sqrt{\frac{\bar b}b}
\Omega_{\bar w}(\bar\beta-b\gamma)-\frac i2\Omega_z(\bar b\bar\alpha+\delta)
+\delta\partial_z\ln\bar b\ ,
\end{align}
\begin{eqnarray}
\partial\alpha&=&-\frac ib(\Omega_w+b\bar b\Omega_z\sigma_w)(b\alpha+\bar
\delta)+2ie^{-\Phi}|b|\Omega_w\sigma_w(\beta-\bar b\bar\gamma)\ ,\nonumber\\
\partial\beta&=&-\frac{ie^{\Phi}}2\sqrt{\frac{\bar b}b}\left(\Omega_z-
4e^{-2\Phi}b\bar b\Omega_{\bar w}\sigma_w+\frac{4\X}{\bar b}\right)(b\alpha+\bar
\delta)-\beta\partial(\Phi-\ln|b|)\nonumber\\
&&+ib\bar b\Omega_z\sigma_w(\beta-\bar b\bar\gamma)+4i(b\X+\bar b\Xb)\sigma_w
\beta-4ib\bar b\X\sigma_w\bar\gamma\ ,\nonumber\\
\partial\gamma&=&\frac ib(\Omega_w+b\bar b\Omega_z\sigma_w)(\bar\beta-b\gamma)
+\gamma\partial\left(\Phi-\frac12\ln\frac b{\bar b}\right)\nonumber\\
&&+2i|b|e^{-\Phi}\Omega_w\sigma_w(\bar b\bar\alpha+\delta)+4i\X\sigma_w
\bar\beta-4i(b\X+\bar b\Xb)\sigma_w\gamma\ ,\nonumber\\
\label{htgr3}\partial\delta&=&ib\bar b\Omega_z\sigma_w(\bar b\bar\alpha+\delta)
+\frac{ie^\Phi}2\sqrt{\frac{\bar b}b}(\Omega_z-4e^{-2\Phi}b\bar b\Omega_{\bar w}
\sigma_w)(\bar\beta-b\gamma)\nonumber\\
&&-2i\X e^\Phi\sqrt{\frac b{\bar b}}\gamma+\delta\partial\ln\bar b\ ,
\end{eqnarray}
\begin{eqnarray}
\bar\partial\alpha&=&-i\bar b\Omega_z\sigma_{\bar w}(b\alpha+\bar\delta)+
\frac{2i\X e^\Phi}{\bar b|b|}\beta
+\frac{ie^\Phi}{2|b|}(\Omega_z+4b\bar b e^{-2\Phi}\Omega_w\sigma_{\bar w})
(\beta-\bar b\bar\gamma)\ ,\nonumber\\
\bar\partial\beta&=&-i(\Omega_{\bar w}-b\bar b\Omega_z\sigma_{\bar w})
(\beta-\bar b\bar\gamma)+\beta\bar\partial(\Phi+\ln|b|)\nonumber\\
&&+2ie^{-\Phi}\bar b|b|\Omega_{\bar w}\sigma_{\bar w}(b\alpha+\bar\delta)+
4i(b\X+\bar b\Xb)\sigma_{\bar w}\beta-4ib\bar b\X\sigma_{\bar w}\bar\gamma\ ,
\nonumber\\
\bar\partial\gamma&=&\frac{ie^\Phi}{2|b|}\left(\Omega_z+4b\bar b e^{-2\Phi}
\Omega_w\sigma_{\bar w}+\frac{4\X}{\bar b}\right)(\bar b\bar\alpha+\delta)-
\gamma\bar\partial\left(\Phi+\frac12\ln\frac b{\bar b}\right)\nonumber\\
&&+i\bar b\Omega_z\sigma_{\bar w}(\bar\beta-b\gamma)+4i\X\sigma_{\bar w}
\bar\beta-4i(b\X+\bar b\Xb)\sigma_{\bar w}\gamma\ ,\nonumber\\
\label{htgr4}\bar\partial\delta&=&-i\left(\Omega_{\bar w}-b\bar b\Omega_z
\sigma_{\bar w}\right)(\bar b\bar\alpha+\delta)-2ie^{-\Phi}\bar b|b|
\Omega_{\bar w}\sigma_{\bar w}(\bar\beta-b\gamma)+\delta\bar\partial\ln\bar b\ ,
\end{eqnarray}
where $\X=X^Ig_I$ and $\Omega_\mu=A_\mu-i\partial_\mu\ln\bar b$.

To proceed it is convenient to set $b=re^{i\varphi}$ and to introduce the new
basis\footnote{Note that the first Killing spinor has components $(1,0,0,0)$
in this basis.}
\eq
\vec\psi=\left(
\begin{array}{c}
\psi_0\\
\psi_1\\
\psi_2\\
\psi_{12}
\end{array}
\right)=\left(
\begin{array}{c}
\alpha\\
-r^2\alpha-\bar b\bar\delta\\
re^{-\Phi}\bar b\bar\gamma\\
re^{-\Phi}\beta
\end{array}
\right)\ , \label{basis-psi}
\feq
in which the gaugini conditions (\ref{htg1})-(\ref{htg4}) become
\begin{align}
\label{htgIII1}\bar\psi_-\partial_zz^\alpha+2e^{-2\Phi}\bar\psi_1\partial
z^\alpha&=-\frac{4i}b g^{\alpha\bar\beta}\mathcal{D}_{\bar\beta}\bar X^Ig_I
\bar\psi_2\ ,\\
\label{htgIII2}\bar\psi_1\partial_zz^\alpha-2\bar\psi_-\bar\partial
z^\alpha&=0\ ,\\
\label{htgIII3}\psi_1\partial_zz^\alpha-2\psi_-\partial z^\alpha&=0\ ,\\
\label{htgIII4}\psi_-\partial_zz^\alpha+2e^{-2\Phi}\psi_1\bar\partial z^\alpha
&=\frac{4i}b g^{\alpha\bar\beta}\mathcal{D}_{\bar\beta}\bar X^Ig_I\psi_{12}\ ,
\end{align}
with $\psi_{\pm}=\psi_2\pm\psi_{12}$.
In general the Killing spinor equations do not readily provide
information and one has to resort to their integrability conditions.
Rewriting the linear system \eqref{htgr1}-\eqref{htgr4} in the basis
\eqref{basis-psi}, and defining $Q=e^{-2\Phi}\bar bD\bar b$, $P=e^{-2\Phi}bDb$,
one finds that the $t$-$w$ integrability condition implies
\begin{align}
\label{httw1} -\frac 12 \left(D_zQ-ie^{-2\Phi}\bar b^2F_{zw}\right)\psi_1
+\left(DQ\right)\psi_-&=0\ ,\\
\label{httw2} -\frac 12 \left(D_zP+ie^{-2\Phi}b^2F_{zw}\right)\psi_1
+\left(DP\right)\psi_-&=0\ ,\\
\label{httw3}f_A\psi_1+f_B\psi_--2i\partial(b\X)\psi_2&=0\ ,\\
\label{httw4}f_C\psi_1+f_D\psi_-+2i\partial(\bar b\Xb)\psi_{12}&=0\ ,
\end{align}
where $F_{\mu\nu}$ denotes the field strength of the K\"ahler U$(1)$ \eqref{KaehlerU(1)}, and
\begin{align}
f_A=&\frac{\bar b}{2b}\left[-2e^{-2\Phi}Db\bar D b+2e^{-2\Phi}bD\bar D b
-(D_zb)^2+6i\Xb D_zb+8(\Xb)^2\right]\ ,\nonumber\\
f_B=&\frac{\bar b}{2b}e^{2\Phi}(D_z P+ie^{-2\Phi}b^2F_{zw})-2i[\X Db+\bar bD\Xb]\ ,
\nonumber\\ 
f_C=&-\frac b{2\bar b}\left[-2e^{-2\Phi}D\bar b\bar D\bar b+2e^{-2\Phi}\bar bD\bar D
\bar b-(D_z\bar b)^2-6i\X D_z\bar b+8(\X)^2\right]\ ,\nonumber\\
f_D=&-\frac b{2\bar b}e^{2\Phi}(D_z Q-ie^{-2\Phi}\bar b^2F_{zw})-2i[\Xb D\bar b+bD\X]\ .\nonumber
\end{align}

\subsection{Time-dependence of second Killing spinor}
\label{time-dep-Kill}

In this subsection we will make use of the Killing spinor equations
\eqref{htgr1}-\eqref{htgr4} and the integrability conditions
(\ref{httw1})-(\ref{httw4}) to derive the time-dependence of the second
Killing spinor. Let us define $\texttt{g}(t,z,w,\bar w)$ by
\[
\psi_-=\frac12\texttt{g}(t,z,w,\bar w)(D_zP+ie^{-2\Phi}b^2F_{zw})\ .
\]
Plugging this into (\ref{httw2}), one gets under the assumption
$D_zP+ie^{-2\Phi}b^2F_{zw}\neq 0$\footnote{The case
$D_zP+ie^{-2\Phi}b^2F_{zw}=0$ will be considered in appendix \ref{DzP}.}
\[
\psi_1=\texttt{g}DP\ .
\]
Using this form of $\psi_-$ and $\psi_1$, the integrability
condition (\ref{httw3}) becomes
\begin{equation}
\label{httw3II}f_A\texttt{g}DP+f_B\frac{\texttt{g}}2
(D_zP+ie^{-2\Phi}b^2F_{zw})-2i\psi_2\partial(b\X)=0\ .
\end{equation}
Now, if $\texttt{g}=0$ the gravitini equations
(\ref{htgr1})-(\ref{htgr4}) imply that $\X=0$. If we exclude for
the time being this degenerate subcase, we have
$\texttt{g}\neq0$ and thus $\texttt{g}=:e^\texttt{G}$.
Dividing (\ref{httw3II}) by $\texttt{g}$ and deriving with respect
to $t$ yields $\partial_t(\psi_2/\texttt{g})=0$ (if $\partial(b\X)\neq 0$)
and hence
\[
\psi_2=e^\texttt{G}\tilde{\psi}_2(z,w,\bar w)\ .
\]
It is then clear that $\partial_t\psi_\texttt{i}=\psi_\texttt{i}\partial_t\texttt{G}$,
$\texttt{i}=1,2,12$. The Killing spinor equations are of the form
$\partial_\mu\psi_\texttt{i}=\mathcal{M}_{\mu\texttt{i}\texttt{j}}\psi_\texttt{j}$,
for some time-independent matrices $\mathcal{M}_\mu$. Taking the
derivative of this with respect to $t$, one gets
$\partial_\mu\partial_t\texttt{G}=0$, and therefore
\[
\texttt{G}=\texttt{G}_0t+\tilde{\texttt{G}}(z,w,\bar w)\ ,
\]
with $\texttt{G}_0\in \mathbb{C}$ constant. We have thus
\eq
\partial_t\psi_{\texttt{i}}=\texttt{G}_0\psi_{\texttt{i}} \label{time-der-psi}
\feq
Furthermore the time-dependence of $\psi_0$ can be easily deduced
from the Killing spinor equations for $\psi_0$,
\begin{align}
\label{httda0}\partial_t\psi_0=&i\Omega_z\psi_1-2i\Omega_w\psi_-\ ,\\
\label{htzda0}\partial_z\psi_0=&\frac i{2r^2}\Omega_z\psi_1+
\frac i{r^2}\Omega_w\psi_-\ ,\\
\label{htwda0}\partial\psi_0=&\left(\frac i{r^2}\Omega_w+i\Omega_z\sigma_w
\right)\psi_1-2i\Omega_w\sigma_w\psi_-\ ,\\
\label{htwbda0}\bar\partial\psi_0=&i\Omega_z\sigma_{\bar w}\psi_1
-\left(\frac{ie^{2\Phi}}{2r^2}\Omega_z+2i\Omega_w\sigma_{\bar w}\right)\psi_-
+\frac{2i\X e^{2\Phi}}{\bar b r^2}\psi_{12}\ .
\end{align}
Deriving \eqref{httda0}-\eqref{htwbda0} with respect to $t$ and taking into
account \eqref{time-der-psi}, one obtains
$\partial_t\partial_\mu\psi_0=\texttt{G}_0\partial_\mu\psi_0$.
Hence $\partial_t\psi_0=\texttt{G}_0\psi_0+\lambda$ where
$\lambda$ is an arbitrary constant. If $\texttt{G}_0\neq 0$, this implies
\eq
\psi_0 = -\frac{\lambda}{\texttt{G}_0} + \tilde\psi_0(z,w,\bar w)
e^{\texttt{G}_0t}\ .
\feq
In that case one can set $\lambda=0$ without loss of generality, because
a nonvanishing $\lambda$ simply corresponds to adding a multiple of the
first Killing spinor to the second. The time-dependence of $\psi_0$ is thus
of the same exponential form as that of the other components of the second
Killing spinor,
\[
\psi_0=\tilde{\psi}_0(z,w,\bar w)e^{\texttt{G}_0t}\ ,\qquad
\psi_{\texttt{i}}=\tilde{\psi}_{\texttt{i}}(z,w,\bar w)e^{\texttt{G}_0t}\ .
\]
If $\texttt{G}_0$ vanishes we have
\eq
\psi_0 = \lambda t + \breve{\psi}_0(z,w,\bar w)\ , \qquad
\psi_{\texttt{i}} = \breve{\psi}_{\texttt{i}}(z,w,\bar w)
\feq
(so that one cannot choose $\lambda=0$ in this case).

Plugging this time-dependence into the subsystem of the Killing
spinor equations not containing $\psi_0$ one obtains the following
reduced system for $\psi_{\texttt{i}}$:
\begin{align}
\label{thpsi01}\partial_z\psi_1&+\left(\frac{\texttt{G}_0}{2b\bar b}-
\frac{\partial_zb}b+iA_z\right)\psi_1+2\left(\frac{\partial b}b-iA_w\right)
\psi_-=0\ ,\\
\label{thpsi02}\partial_z\psi_2&+\left(\frac{\texttt{G}_0}{2b\bar b}-
\frac{\partial_z\bar b}{\bar b}-4i\frac{\X}{\bar b}-iA_z\right)\psi_2
-\left(\frac{\partial_zb}b-4i\frac{\Xb}b-iA_z\right)\psi_{12}=0\ ,\\
\label{thpsi03}\partial_z\psi_{12}&+2e^{-2\Phi}\left(\frac{\bar\partial\bar b}
{\bar b}+iA_{\bar w}\right)\psi_1
+\left(\frac{\texttt{G}_0}{2b\bar b}-\frac{\partial_zb}b-\frac{\partial_z
\bar b}{\bar b}-4i\frac{\X}{\bar b}\right)\psi_{12}=0\ ,
\end{align}
\begin{align}
\label{thpsi04}\partial_z\psi_1&-\left(\frac{\texttt{G}_0}{2b\bar b}+
\frac{\partial_z\bar b}{\bar b}+iA_z\right)\psi_1
+2\left(\frac{\partial\bar b}{\bar b}+iA_w\right)\psi_-=0\ ,\\
\label{thpsi05}\partial_z\psi_2&-2e^{-2\Phi}\left(\frac{\bar\partial b}b
-iA_{\bar w}\right)\psi_1
-\left(\frac{\texttt{G}_0}{2b\bar b}+\frac{\partial_zb}b+\frac{\partial_z
\bar b}{\bar b}-4i\frac{\Xb}b\right)\psi_2=0\ ,\\
\label{thpsi06}\partial_z\psi_{12}&-\left(\frac{\partial_z\bar b}{\bar b}
+4i\frac{\X}{\bar b}+iA_z\right)\psi_2
-\left(\frac{\texttt{G}_0}{2b\bar b}+\frac{\partial_zb}b-4i\frac{\Xb}b-iA_z
\right)\psi_{12}=0\ ,
\end{align}
\begin{align}
\label{thpsi07}\partial\psi_1&-\texttt{G}_0\sigma_w\psi_1=0\ ,\\
\label{thpsi08}\partial\psi_2&+\left(\frac{\partial_zb}{2b}-2i\frac{\Xb}b-
\frac i2A_z\right)\psi_1
-\left(\texttt{G}_0\sigma_w+\frac{\partial b}b+\frac{\partial\bar b}{\bar b}
-2\partial\Phi\right)\psi_2=0\ ,\\
\label{thpsi09}\partial\psi_{12}&-\left(\frac{\partial_z\bar b}{2\bar b}+
2i\frac{\X}{\bar b}+\frac i2A_z\right)\psi_1
-\left(\texttt{G}_0\sigma_w+\frac{\partial b}b+\frac{\partial\bar b}{\bar b}
-2\partial\Phi\right)\psi_{12}=0\ ,
\end{align}
\begin{align}
\label{thpsi10}\bar\partial\psi_1&-\left(\texttt{G}_0\sigma_{\bar w}+
\frac{\bar\partial b}b+\frac{\bar\partial\bar b}{\bar b}\right)\psi_1\nonumber\\
&-e^{2\Phi}\left[\left(\frac{\partial_zb}{2b}+\frac{\partial_z\bar b}{2\bar b}
\right)\psi_--2i\left(\frac{\Xb}b\psi_2+\frac{\X}{\bar b}\psi_{12}\right)
\right]=0\ ,\\
\label{thpsi11}\bar\partial\psi_2&-\left(\texttt{G}_0\sigma_{\bar w}+
\frac{\bar\partial\bar b}{\bar b}+iA_{\bar w}\right)\psi_2
-\left(\frac{\bar\partial b}b-iA_{\bar w}\right)\psi_{12}=0\ ,\\
\label{thpsi12}\bar\partial\psi_{12}&-\left(\frac{\bar\partial\bar b}
{\bar b}+iA_{\bar w}\right)\psi_2-\left(\texttt{G}_0\sigma_{\bar w}+
\frac{\bar\partial b}b-iA_{\bar w}\right)\psi_{12}=0\ .
\end{align}
From the difference of eqns.~(\ref{thpsi02})-(\ref{thpsi06}) and
(\ref{thpsi11})-(\ref{thpsi12}) one gets respectively
\begin{equation}
\partial_z\psi_-=-\frac{\texttt{G}_0}{2b\bar b}\psi_+\ ,\qquad
\bar\partial\psi_-=\texttt{G}_0\sigma_{\bar w}\psi_-\ . \label{deriv-psi_-}
\end{equation}
Furthermore, $[(\ref{thpsi05})-(\ref{thpsi03})-2e^{-2\Phi}(\ref{thpsi10})]$
yields
\begin{equation}
\bar\partial\psi_1=\frac{e^{2\Phi}}2\partial_z\psi_--\texttt{G}_0
\left(\frac{e^{2\Phi}}{4b\bar b}\psi_+-\sigma_{\bar w}\psi_1\right)\ .
\label{deriv-psi_1}
\end{equation}
Obviously for $\texttt{G}_0=0$, the equations (\ref{thpsi01})-(\ref{thpsi12})
simplify significantly. Let us now study this particular case under
the additional assumption $\psi_-\neq0$ and $\psi_1\neq0$.

\subsection {Case $\texttt{G}_0=0$, $\psi_-\neq0$ and $\psi_1\neq0$}
\label{psi_-neq0}

For $\texttt{G}_0=0$ one gets from \eqref{thpsi07}, \eqref{deriv-psi_-} and
\eqref{deriv-psi_1}
\[
\psi_1=\psi_1(z)\ ,\qquad \psi_-=\psi_-(w)\ .
\]
Assuming $\psi_-\neq0$, the gaugini equations \eqref{htgIII1}-\eqref{htgIII4}
imply
\begin{eqnarray}
\label{htgIV1}\partial_zz^\alpha&=&-\frac{4i}b g^{\alpha\bar\beta}
\mathcal{D}_{\bar\beta}\bar X^Ig_I\frac{\psi_-\bar\psi_2}{\psi_-\bar\psi_-+
e^{-2\Phi}\psi_1\bar\psi_1}\ ,\\
\label{htgIV2}\partial z^\alpha&=&\frac{\psi_1}{2\psi_-}\partial_zz^\alpha\ ,\\
\label{htgIV3}\bar\partial z^\alpha&=&\frac{\bar\psi_1}{2\bar\psi_-}\partial_z
z^\alpha\ ,\\
\label{htgIV4}0&=&g^{\alpha\bar\beta}\mathcal{D}_{\bar\beta}\bar
X^Ig_I\left(\psi_2\bar\psi_2-\psi_{12}\bar\psi_{12}\right)\ .
\end{eqnarray}
From eqns.~(\ref{htgIV2}) and (\ref{htgIV3}) we obtain
\begin{equation}
A_z\psi_1-2A_w\psi_-=0\ . \label{AzAw}
\end{equation}
(\ref{thpsi01})$+$(\ref{thpsi04}) and (\ref{thpsi03})$-$(\ref{thpsi05}) yield
respectively
\begin{align}
\label{thG01}&\partial_z\psi_1=\psi_1\partial_z\ln|b|-2\psi_-\partial\ln|b|\ ,\\
\label{thG02}&0=\psi_-\partial_z\ln|b|+2e^{-2\Phi}\psi_1\bar\partial\ln|b|
-2i\left(\frac{\Xb}b\psi_2+\frac{\X}{\bar b}\psi_{12}\right)\ .
\end{align}
Using (\ref{thG01}) and (\ref{thG02}) it is easy to shew that
\begin{equation}
\label{thG03}\bar\psi_1\partial_z\psi_1-\psi_1\partial_z\bar\psi_1=2ie^{2\Phi}
\left(\frac{\X}{\bar b}+\frac{\Xb}b\right)(\psi_2\bar\psi_2-
\psi_{12}\bar\psi_{12})\ .
\end{equation}
Because we are interested only in the case in which
$g^{\alpha\bar\beta}\mathcal{D}_{\bar\beta}\bar X^Ig_I\neq0$\footnote{One readily
shows that $g^{\alpha\bar\beta}\mathcal{D}_{\bar\beta}\bar X^Ig_I=0$ leads to
$\partial_{\bar\beta}V=0$, where $V$ is the scalar potential \eqref{scal-pot}.
Unless there are flat directions in the potential, these equations
completely fix the moduli which are thus constant.}, (\ref{htgIV4}) implies
$|\psi_2|=|\psi_{12}|$ and thus from (\ref{thG03}) one gets
\begin{equation}
\label{thG04}\bar\psi_1\partial_z\psi_1-\psi_1\partial_z\bar\psi_1=0\ .
\end{equation}
Hence $\psi_1=\zeta(z)e^{i\theta_0}$ where $\theta_0$ is a constant
and $\zeta(z)$ is a real function. By rescaling
$\psi_{\texttt{i}}\rightarrow e^{-i\theta_0}\psi_{\texttt{i}}$ we can take
$\psi_1$ real and positive without loss of generality. By assumption both
$\psi_1$ and $\psi_-$ are non-vanishing, which allows to introduce new
coordinates $Z$, $W$ and $\bar W$ such that
\[
dZ=-\frac{2dz}{\psi_1(z)}\ ,\qquad dW=\frac{dw}{\psi_-(w)}\ ,\qquad d\bar
W=\frac{d\bar w}{\bar\psi_-(\bar w)}\ .
\]
Note that one can set $\psi_-=1$ using the residual gauge invariance
$w\mapsto W(w)$,
$\Phi\mapsto\Phi-\frac12\ln(dW/dw)-\frac12\ln(d\bar W/d\bar w)$ leaving
invariant the metric $e^{2\Phi}dwd\bar w$. We can thus take $W=w$ in the
following. (\ref{thpsi01}) and (\ref{thpsi04}) are then equivalent to
\[
(\partial_Z+\partial)\varphi=0\ ,\qquad
\partial_Z\ln\psi_1-(\partial_Z+\partial)\ln r=0\ .
\]
From the real part of the first equation one has
\[
\varphi=\varphi(Z-w-\bar w)\ .
\]
Using $\psi_1=\psi_1(Z)$, the second equation implies
\begin{equation}
(\partial_Z+\partial)\frac{r}{\psi_1}=0\ , \label{r/psi_1}
\end{equation}
and therefore
\[
\frac{r}{\psi_1}=\rho(Z-w-\bar w)\ .
\]
The function $b$ must thus have the form
\[
b(Z,w,\bar w)=\psi_1(Z)B(Z-w-\bar w)\ ,
\]
where $B(Z-w-\bar w)=\rho(Z-w-\bar w)e^{i\varphi(Z-w-\bar w)}$. Taking into
account \eqref{dzPhi} and \eqref{r/psi_1}, the difference between
(\ref{thpsi08}) and (\ref{thpsi09}) yields
\[
(\partial_Z+\partial)(\ln\psi_1-\Phi)=0\ ,
\]
so that $\ln\psi_1-\Phi=-H(Z-w-\bar w)$ with $H$ real. This gives
\[
e^{2\Phi}=\psi_1^2e^{2H}
\]
for the conformal factor. The conditions (\ref{htgIV1})-(\ref{htgIV4})
coming from the gaugino variations boil down to
\begin{align}
\label{htgV1}&\partial_Z
z^\alpha=\frac iBg^{\alpha\bar\beta}\mathcal{D}_{\bar\beta}\bar X^Ig_I
\frac{1-\psi_+}{1+e^{-2H}}\ ,\\
\label{htgV2}&\partial z^\alpha=\bar\partial z^\alpha=-\partial_Z z^\alpha\ ,\\
\label{htgV3}&\bar\psi_+=-\psi_+\ .
\end{align}
From equation (\ref{htgV2}) we obtain that $z^{\alpha}=z^{\alpha}(Z-w-\bar w)$.
In terms of the new coordinate $Z$, \eqref{dzPhi} reads
\[
\partial_Z\Phi+i\left(\frac{\Xb}{B}-\frac{\X}{\bar B}\right)=0\ .
\]
Using the definition of $H$ we get
\begin{equation}
\label{thpsic}\partial_Z\ln\psi_1=-\dot{H}-i\left(\frac{\Xb}B-\frac{\X}{\bar B}
\right)\ ,
\end{equation}
where a dot denotes a derivative w.r.t.~$Z-w-\bar w$. As the lhs depends only
on $Z$ and the rhs depends only on $Z-w-\bar w$, we can conclude that
$\partial_Z\ln\psi_1=\kappa$ with some real constant $\kappa$, i.e.,
$\psi_1(Z)=\psi_1^{(0)}e^{\kappa Z}$. By shifting $Z$ one can set
$\psi_1^{(0)}=1$. The only remaining nontrivial equations in the system
(\ref{thpsi01})-(\ref{thpsi12}) read
\begin{align}
\label{thpsiII1}&\partial_Z\psi_+-2\left(\frac{\dot{\rho}}{\rho}-\dot{H}\right)
\psi_++2i\left(\dot{\varphi}-A_Z\right)+2i\left(\frac{\Xb}B+\frac{\X}{\bar B}
\right)=0\ ,\\
\label{thpsiII2}&\partial_Z\psi_+-\left(2\frac{\dot{\rho}}{\rho}-\dot{H}+
\kappa\right)\psi_+-2ie^{-2H}\left(\dot{\varphi}-A_Z\right)-i\left(\frac{\Xb}B
+\frac{\X}{\bar B}\right)=0\ ,\\
\label{thpsiII3}&\partial\psi_++2\left(\frac{\dot{\rho}}{\rho}-\dot{H}\right)
\psi_+-2i\left(\dot{\varphi}-A_Z\right)-2i\left(\frac{\Xb}B+\frac{\X}{\bar B}
\right)=0\ ,\\
\label{thpsiII4}&\bar\partial\psi_++2\frac{\dot{\rho}}{\rho}\psi_+-
2i\left(\dot{\varphi}-A_Z\right)=0\ ,\\
\label{thpsiII5}&i\left(\frac{\Xb}B+\frac{\X}{\bar B}\right)\psi_++
2\left(1+e^{-2H}\right)\frac{\dot{\rho}}{\rho}-\dot{H}+\kappa=0\ .
\end{align}
From (\ref{thpsiII1})+(\ref{thpsiII3}) and
(\ref{thpsiII1})+(\ref{thpsiII4}) we obtain respectively
\begin{align}
\label{thpsiIIc1}&\left(\partial_Z+\partial\right)\psi_+=0\ ,\\
\label{thpsiIIc2}&\left(\partial_Z+\bar\partial\right)\psi_+=-2\dot{H}\psi_+
-2i\left(\frac{\Xb}B+\frac{\X}{\bar B}\right)\ .
\end{align}
Since $\psi_+$ is imaginary (cf.~\eqref{htgV3}), \eqref{thpsiIIc1} implies
$\psi_+=\psi_+(Z-w-\bar w)$ so that \eqref{thpsiIIc2} yields
\begin{equation}
\label{thpsiIIc3}\dot{H}\psi_++i\left(\frac{\Xb}B+\frac{\X}{\bar B}\right)=0\ .
\end{equation}
Using these informations, eqns.~(\ref{thpsiII1})-(\ref{thpsiII5}) reduce
further to
\begin{align}
\label{thpsiIII1}\left[\left(1+e^{2H}\right)\frac{\psi_+}{\rho^2}
\right]^{\!\text{\large{$\cdot$}}}-\kappa e^{2H}\frac{\psi_+}{\rho^2}&=0\ ,\\
\label{thpsiIII2}\left(\frac{\psi_+}{\rho^2}\right)^{\!\text{\large{$\cdot$}}}
+2i\frac{\dot{\varphi}-A_Z}{\rho^2}&=0\ ,\\
\label{thpsiIII3}\dot{H}\left(1+\psi_+^2\right)-2\frac{\dot{\rho}}{\rho}
\left(1+e^{-2H}\right)&=\kappa\ .
\end{align}
Eliminating $\dot{\rho}/\rho$ from \eqref{thpsiIII1} and \eqref{thpsiIII3}
leads to
\begin{equation}
\label{thpsiIV1-3b}\dot{H}\psi_+(1-\psi_+^2)+(1+e^{-2H})\dot{\psi}_+=0\ ,
\end{equation}
that can be integrated to give
\begin{equation}
\label{thpsi+}\psi_+=\frac{ia}{\sqrt{1+e^{2H}-a^2}}\ ,
\end{equation}
where $a$ is real integration constant. To proceed we observe that from
\eqref{thpsic} and \eqref{thpsiIIc3} one obtains for the function $B$,
\begin{equation}
\label{thB} B=-\frac{2i\Xb}{\dot H(1+\psi_+)+\kappa}\ ,
\end{equation}
and thus for its absolute value $\rho$ and phase $\varphi$
\begin{align}
\label{thBph}\rho^{-2}&=\frac{(\kappa+\dot{H})^2-\dot{H}^2\psi_+^2}{4\X\Xb}\ ,\\
\label{thBmo}\tan\varphi&=i\frac{(\X+\Xb)(\kappa+\dot{H})+(\X-\Xb)(\dot{H}
\psi_+)}{(\X-\Xb)(\kappa+\dot{H})+(\X+\Xb)(\dot{H}\psi_+)}\ .
\end{align}
Using \eqref{thBph}, \eqref{thpsiIII3} yields a relation between $H$ and $\X$,
\begin{eqnarray}
\label{thpsiIV1+3}0&=&2\left(1+e^{-2H}\right)\ddot{H}+\dot{H}^2\left(1+3\psi_+^2\right)-\kappa^2\nonumber\\
&&-\frac{\left(\dot{H}+\kappa\right)^2-\dot{H}^2\psi_+^2}{\dot{H}\left(1-\psi_+^2\right)+\kappa}
\left(1+e^{-2H}\right)\left(\frac{\Xp}{\X}+\frac{\Xbp}{\Xb}\right)\ ,
\end{eqnarray}
while (\ref{thpsiIII2}) gives $A_Z$,
\begin{align}\label{thH1}
&A_Z=\frac i2\left\{(1+\psi_+)\frac{\Xp}{\X}-(1-\psi_+)\frac{\Xbp}{\Xb}\right. \\
&\left.-\frac{\dot{H}\psi_+\left(1-\psi_+^2\right)\left(1+e^{-2H}\right)^{-1}}{\left(\dot{H}+\kappa\right)^2-\dot{H}^2\psi_+^2}\left[2\left(1+e^{-2H}\right)\ddot{H}+\dot{H}^2\left(1+3\psi_+^2\right)-\kappa^2\right]\right\}\ . \nonumber
\end{align}
Making use of \eqref{thpsiIV1+3}, this boils down to
\begin{equation}\label{thAZ}
A_Z=-\left[\dot{H}\left(1-\psi_+^2\right)+\kappa\right]^{-1}
\mbox{Im}\left\{\left[\dot{H}\left(1-\psi_+\right)+\kappa\right](1+\psi_+)\frac{\Xp}{\X}\right\}\ .
\end{equation}
The condition \eqref{Delta-Phi} is then automatically satisfied: Plugging the relation
\begin{displaymath}
\Xp+iA_Z\X=\dot{z}^{\alpha}{\cal D}_{\alpha}\X=\frac iB g^{\alpha\bar\beta}{\cal D}_{\alpha}\X
{\cal D}_{\bar\beta}\Xb\frac{1-\psi_+}{1+e^{-2H}}\ ,
\end{displaymath}
(where we used \eqref{htgV1} in the second step) into
\begin{displaymath}
-\frac12(\text{Im}\,{\cal N})^{-1|IJ}g_Ig_J=\X\Xb+g^{\alpha\bar\beta}{\cal D}_{\alpha}\X
{\cal D}_{\bar\beta}\Xb\ ,
\end{displaymath}
that follows from special geometry \cite{Vambroes}, one gets
\begin{displaymath}
(\mbox{Im}\,\mathcal{N})^{-1|IJ}g_Ig_J=-2\X\Xb+\frac{4\Xb}{\dot{H}(1+\psi_+)+\kappa}
\frac{1+e^{-2H}}{1-\psi_+}\left(\Xp+iA_Z\X\right)\ .
\end{displaymath}
Inserting this into \eqref{Delta-Phi}, the latter becomes
\begin{eqnarray}
\label{thF=dA}0&=&2\left(1+e^{-2H}\right)\ddot{H}+\dot{H}^2\left(1+3\psi_+^2\right)-\kappa^2\nonumber\\
&&-2\left[\dot{H}(1-\psi_+)+\kappa\right]\frac{1+e^{-2H}}{1-\psi_+}
\left(\frac{\Xp}{\X}+i A_Z\right)\ ,
\end{eqnarray}
which coincides with (\ref{thpsiIV1+3}) once we substitute in it the
expression (\ref{thAZ}) for $A_Z$.

The Bianchi identities (\ref{bianchi}) and Maxwell equations \eqref{maxwell}
can be integrated once, with the result
\begin{eqnarray}\label{thbianchi}
&&(1+e^{2H})\left(\frac{X^I}{\bar B}-\frac{\bar
X^I}B\right)^{\!\text{\large{$\cdot$}}}-\kappa e^{2H} \left(\frac{X^I}{\bar
B}-\frac{\bar X^I}B\right)\nonumber\\
&+&ie^{2H}\left[\frac{\left(\mbox{Im}\,\mathcal{N}\right)^{-1|IJ}g_J}{B\bar
B}+2i\dot{H}\psi_+\left(\frac{X^I}{\bar B}+\frac{\bar X^I}B\right)\right]=ip^I\ ,
\end{eqnarray}
\begin{eqnarray}\label{thmaxwell}
&&(1+e^{2H})\left(\frac{F_I}{\bar B}-\frac{\bar
F_I}B\right)^{\!\text{\large{$\cdot$}}}-\kappa e^{2H} \left(\frac{F_I}{\bar
B}-\frac{\bar F_I}B\right)-g_Ie^{2H}\frac{\psi_+}{\rho^2}\nonumber\\
&+&ie^{2H}\left[\frac{\mbox{Re}\,\mathcal{N}_{IL}\left(\mbox{Im}\,\mathcal{N}\right)^{-1|JL}g_J}{B\bar
B}+2i\dot{H}\psi_+\left(\frac{F_I}{\bar B}+\frac{\bar F_I}B\right)\right]=iq_I\ ,
\end{eqnarray}
where $p^I,q_I$ are integration constants. It is straightforward to show that \eqref{thbianchi}
and \eqref{thmaxwell} are implied by \eqref{htgV1}, \eqref{thpsiIII2}-\eqref{thpsiIV1-3b} and
\eqref{thB} iff $p^I=q_I=0$\footnote{This does not mean that all the fluxes vanish.}.

Finally, the shift vector $\sigma$ follows from \eqref{dsigma} that simplifies to
\eq
\label{thsigma}\partial_Z\sigma_w=\frac{e^{-\kappa
Z}}4\left(\frac{\psi_+}{\rho^2}\right)^{\!\text{\large{$\cdot$}}}\ , \qquad
\partial\sigma_{\bar w}-\bar\partial\sigma_w=-\frac{e^{-\kappa
Z}}2\left(e^{2H}\frac{\psi_+}{\rho^2}\right)^{\!\text{\large{$\cdot$}}}\ ,
\feq
whose solution is
\eq{\label{thsigmaint}}
\sigma=-\frac{e^{-\kappa Z}}4e^{2H}\frac{\psi_+}{\rho^2}(dw-d\bar w)\ .
\feq
Note that in the case $\kappa\neq0$ one can always set $\kappa=1$ by
rescaling the coordinates.

The missing component $\psi_0$ of the second Killing spinor is determined
by the system \eqref{httda0}-\eqref{htwbda0} that can be integrated straightforwardly.
This yields (after going back to the original basis)
\begin{align}
\alpha&=\hat\alpha-2\kappa t-\frac{\psi_1}{2b\bar b}-\frac{e^{2\Phi}\psi_+}{2\psi_1b\bar b}\ , \qquad
\beta=-\frac{e^{\Phi}}{2|b|}\left(1-\psi_+\right)\ , \nonumber\\
\gamma&=-\frac{\beta}b\ , \qquad \delta=-\bar b\bar\alpha-\frac{\psi_1}b
\end{align}
for the second Killing spinor. Here, $\hat\alpha$ denotes an integration constant.
As is clear from \eqref{delta-psi} and \eqref{delta-lambda}, $C(\epsilon^1,\epsilon_2)$,
with $C\in\bC$ an arbitrary constant, is again Killing if $(\epsilon^1,\epsilon_2)$ is.
This means that multiplication of $\alpha$ and $\beta$ by $C$ and of $\gamma$ and $\delta$
by $\bar C$ gives again a solution of the Killing spinor equations.
Choosing $\hat\alpha=1/C$, in order to obtain the first Killing spinor when
$C\rightarrow0$, the norm squared of the associated Killing vector
$V_{\mu}=A(\epsilon^i,\gamma_{\mu}\epsilon_i)$ (with $A$ given in \eqref{Majorana})
turns out to be
\begin{align}
V^2=&-4|b|^2\left\{|1-2\kappa C t|^2-\left[\frac{|C|\psi_1\left(1+e^{2H}\right)}{2|b|^2}\right]^2
\frac{1-a^2}{1+e^{2H}-a^2}\right.\nonumber\\
&\left.+\frac{\psi_1e^{2H}}{|b|^2}\frac{a\mbox{Im}C}{\sqrt{1+e^{2H}-a^2}}\right\}^2
-\left(\frac{2\psi_1\mbox{Im}C}{|b|}\right)^2\ .
\end{align}
For $V^2=0$ the solution belongs also to the null class considered
in \cite{Klemm:2009uw}. This happens for $\mbox{Im}C=0$, $\kappa=0$,
$a^2<1$ and
\begin{equation}\label{htV2=0}
\dot{H}=\sqrt{\frac{8\X\Xb}{|C|(1-a^2)^{1/2}}}\frac{(1+e^{2H}-a^2)^{3/4}}
{1+e^{2H}}\ .
\end{equation}
(\ref{htV2=0}) is actually the general form of $\dot{H}$ in the case $\kappa=0$.
To see this, observe that (\ref{thpsiIII1}) implies
\begin{equation}
(1+e^{2H})\frac{\psi_+}{\rho^2}=ih_0\ ,
\end{equation}
if $\kappa=0$, where $h_0$ is a real integration constant. Using the expressions
(\ref{thpsi+}) and (\ref{thBph}) for $\psi_+$ and $\rho^2$ we obtain
exactly (\ref{htV2=0}), with $h_0$ and $C$ related by
$h_0|C|(1-a^2)^{1/2}=2a$. Plugging the expression for $\dot H$ into
(\ref{htgV1}) we find that the scalars have to satisfy the flow equation
\begin{equation}\label{thg0z}
\dot{z}^\alpha=-\left(\frac{h_0\X}{a\Xb}\right)^{1/2}
\frac{g^{\alpha\bar\beta}\mathcal{D}_{\bar\beta}\Xb}{\left(1+e^{-2H}\right)
\left(1+e^{-2H}-a^2\right)^{1/4}}\ .
\end{equation}
Using $w=x+iy$ and $dZ=\frac{dH}{\dot H}+2dx$, the metric reads
\eq
ds^2 = -4\rho^2\left[dt-e^{2H}\frac{i\psi_+}{2\rho^2}dy\right]^2
+ \frac 1{4\rho^2}\left(\frac{dH}{\dot H}+2dx\right)^2 + \frac{e^{2H}}{\rho^2}
(dx^2+dy^2)\ , \label{metr-kappa=0}
\feq
where $\psi_+$, $\rho^2$ and $\dot H$ are given by \eqref{thpsi+},
\eqref{thBph} and \eqref{htV2=0} respectively. As a check, let us show
that this solution does indeed coincide with one of the 1/2 BPS lightlike
case classified in \cite{Klemm:2009uw}. To this end, consider the
coordinate transformation
\begin{displaymath}
u = \frac{2a}{h_0}(1-a^2)^{-1/2}t + x + \mu(\chi)\ , \qquad
v = \frac t{\sqrt 2} - \frac{h_0}{2\sqrt 2a}(1-a^2)^{1/2}x +
\nu(\chi)\ ,
\end{displaymath}
\begin{displaymath}
\Psi = 4a\left(\frac a{h_0}\right)^{1/2}(1-a^2)^{-1/4}t
- 2\left(\frac{h_0}a\right)^{1/2}(1-a^2)^{3/4}y\ ,
\end{displaymath}
\begin{displaymath}
\coth\chi = (1-a^2)^{-1/2}(1+e^{2H}-a^2)^{1/2}\ ,
\end{displaymath}
with
\begin{displaymath}
\frac{d\nu}{d\chi} = \frac{(\tanh\chi)^{1/2}}{8\sqrt 2(\X\Xb)^{1/2}
(1-a^2)^{1/4}}\left(\frac{h_0}a\right)^{1/2}\ , \qquad
\frac{d\mu}{d\chi} = -\frac{2\sqrt 2a}{h_0}(1-a^2)^{-1/2}
\frac{d\nu}{d\chi}\ .
\end{displaymath}
Then, the metric \eqref{metr-kappa=0}, the fluxes \eqref{fluxes} and the flow
equation \eqref{thg0z} become
\eq
ds^2 = -2\sqrt2\coth\chi dudv + \frac{d\chi^2}{16\sinh^2\!\chi\X\Xb} +
\frac{d\Psi^2}{2\sinh2\chi}\ ,
\feq
\eq
F^I = \frac{(\text{Im}\,{\cal N})^{-1|IJ}g_J}{4\cosh^2\!\chi(\X\Xb
\tanh\chi)^{1/2}}d\Psi\wedge d\chi\ , \qquad
\frac{dz^{\alpha}}{d\chi} = \frac{g^{\alpha\bar\beta}{\cal D}_{\bar\beta}\Xb}
{\Xb\sinh 2\chi}\ , \label{flow-kappa=0}
\feq
which are exactly the eqns.~(5.33), (5.34) and (5.24) of \cite{Klemm:2009uw}.
We also see that in this case, $a$ can be eliminated by a diffeomorphism,
and thus is not really a parameter of the solution.

\subsubsection{Summary}
\label{summary}

In the case $D_zP+ie^{-2\Phi}b^2F_{zw}\neq0$ and $\texttt{G}_0=0$ and under
the additional assumptions $\psi_-\neq0$ and $\psi_1\neq0$, the
fields are given in terms of the solutions of the system
\eq
\dot{z}^\alpha=-\left[\dot{H}(1+\psi_+)+\kappa\right]\frac{1-\psi_+}{1+e^{-2H}}
\frac{g^{\alpha\bar\beta}\mathcal{D}_{\bar\beta}\Xb}{2\Xb} \label{flow-gen}
\feq
and \eqref{thpsiIV1+3}, where $\kappa=0,1$, the scalars $z^\alpha$ and the
real function $H$ depend only on the combination $Z-w-\bar w$, and $\psi_+$ is
given by \eqref{thpsi+}, with $a\in\mathbb{R}$ an arbitrary constant.
Furthermore, a dot denotes a derivative w.r.t.~$Z-w-\bar w$. Once a solution
($z^{\alpha},H$) is determined, one defines $\rho$ by \eqref{thBph}.
Then, the metric and the fluxes read respectively
\eq
ds^2=-4\rho^2e^{2\kappa Z}\left[dt-e^{2H-\kappa Z}\frac{\psi_+}{4\rho^2}
(dw-d\bar w)\right]^2+\frac1{\rho^2}\left(\frac{dZ^2}4+e^{2H}dwd\bar w\right)\ ,
\feq
\begin{align}
F^I=&8\kappa e^{\kappa Z}\mbox{Im}\left[\frac{\Xb
X^I}{\dot{H}(1+\psi_+)+\kappa}\right]dt\wedge dZ\nonumber\\
&+\frac{2ie^{\kappa Z}}{1+e^{-2H}}\left\{\psi_+\left(\mbox{Im}\mathcal{N}
\right)^{-1|IJ}g_J\right.\nonumber\\
&\left.+4i\kappa\mbox{Im}\left[\frac{(1+\psi_+)\Xb
X^I}{\dot{H}(1+\psi_+)+\kappa}\right]\right\}dt\wedge d(Z-w-\bar w)\nonumber\\
&+\frac{i\left[\left(\dot{H}+\kappa\right)^2-\dot{H}^2\psi_+^2\right]
\left(1+e^{2H}\psi_+^2\right)}{4\X\Xb\left(1+e^{-2H}\right)}
\left\{\left(\mbox{Im}\mathcal{N}\right)^{-1|IJ}g_J\right.\nonumber\\
&\left.+4\kappa\mbox{Re}\left[\frac{\Xb X^I}{\dot{H}(1+\psi_+)+\kappa}\right]
\right\}\left[\frac{dZ}2\wedge(dw-d\bar w)+dw\wedge d\bar w\right]\ .
\end{align}

\subsubsection{Explicit solutions}

We shall now give some explicit solutions for the simple model determined by
the prepotential $F=-iZ^0Z^1$ that has $n_V=1$ (one vector multiplet), and thus
just one complex scalar $\tau$. Choosing $Z^0=1$, $Z^1=\tau$
(cf.~\cite{Vambroes}), the symplectic vector $v$ reads
\eq
v = \left(\begin{array}{c} 1 \\ \tau \\ -i\tau \\ -i\end{array}\right)\ .
\feq
The K\"ahler potential, metric and kinetic matrix for the vectors
are given respectively by
\eq e^{-{\cal K}} = 2(\tau + \bar\tau)\ ,
\qquad g_{\tau\bar\tau} = \partial_\tau\partial_{\bar\tau}{\cal K} = (\tau +
\bar\tau)^{-2}\ ,
\feq
\eq
{\cal N} = \left(\begin{array}{cc} -i\tau & 0 \\ 0 & -\frac i\tau
\end{array}\right)\ .
\feq
Note that positivity of the kinetic terms in the action requires
${\mathrm{Re}}\tau>0$. For the scalar potential one obtains
\eq
V = -\frac4{\tau+\bar\tau}(g_0^2 + 2g_0g_1\tau + 2g_0g_1\bar\tau +
g_1^2\tau\bar\tau)\ ,
\feq
which has an extremum at $\tau=\bar\tau=|g_0/g_1|$. In what
follows we assume $g_I>0$. The K\"ahler U(1) is
\eq A_{\mu} = \frac i{2(\tau+\bar\tau)}\partial_{\mu}(\tau-\bar\tau)\ .
\feq
In order to proceed we shall take $\tau=\bar\tau$ (this includes the
extremum of the potential and thus the AdS vacuum). Then $A=0$ and
equation (\ref{thAZ}) imposes $\kappa\psi_+=0$ if $\Xp\neq0$.
The case $\kappa=0$ was considered in generality above, and an explicit
solution of the flow equation \eqref{flow-kappa=0} for the prepotential of this
paragraph can be found in section 4.5 of \cite{Klemm:2009uw} (put
${\cal G}=0$ there). Thus, we shall focus on the case $\psi_+=0$ in the
following. Then, eqns.~(\ref{thpsiIV1+3}) and (\ref{flow-gen}) boil down to
\eq
\label{thscalarIp}2(1+e^{-2H})\ddot{H}+\dot{H}^2-\kappa^2+\left(1+e^{-2H}\right)
(\dot{H}+\kappa)\frac{g_0-g_1\tau}{g_0+g_1\tau}\frac{\dot{\tau}}{\tau}=0\ ,
\feq
\eq
\label{thscalarIIp}\frac{\dot{\tau}}{\tau}=\frac{\dot{H}+\kappa}{1+e^{-2H}}
\frac{g_0-g_1\tau}{g_0+g_1\tau}\ .
\feq
Plugging \eqref{thscalarIIp} into \eqref{thscalarIp} yields an expression
for $\tau$ in terms of $H$ and its derivatives. Reinserting this into
\eqref{thscalarIIp} gives a third order differential equation for $H$ only,
\eq
\left(1+e^{-2H}\right)^2\dddot{H}+\left[\left(3-2e^{-2H}\right)\left(1+e^{-2H}
\right)\ddot{H}+\dot{H}^2-\kappa^2\right]\dot{H}=0\ ,
\feq
that can be integrated twice, with the result
\eq
\dot{H}=\frac1{\left(1+e^{2H}\right)^{1/4}}\sqrt{2E_1+\frac{E_2}{2\left(1+
e^{2H}\right)^{1/2}}+\kappa^2\left(1+e^{2H}\right)^{1/2}}\ ,
\feq
where $E_1$ and $E_2$ are two integration constants. If ${\dot H}\neq0$,
we can use the function $H$ in place of $w+\bar w$ as a new coordinate.
Using $w=x+iy$, in the coordinate system $\{t,H,y,Z\}$ the solution is given by
\begin{align}
ds^2=&-\left[\frac{2(g_0+g_1\tau)}{\sqrt\tau\left(\dot{H}+\kappa\right)}
\right]^2e^{2\kappa Z}dt^2\nonumber\\
&+\left[\frac{2(g_0+g_1\tau)}{\sqrt\tau\left(\dot{H}+\kappa\right)}\right]^{-2}
\left[dZ^2+e^{2H}\left(dZ-\frac{dH}{\dot{H}}\right)^2+4e^{2H}dy^2\right]\ ,
\label{metr-expl}
\end{align}
\begin{align}
F^0=&-\frac{\left(\dot{H}+\kappa\right)\left(\kappa g_1\tau-g_0\dot{H}\right)}
{\dot{H}\left(g_0+g_1\tau\right)^2\left(1+e^{-2H}\right)}dH\wedge dy\ ,
\nonumber\\
F^1=&-\frac{\tau\left(\dot{H}+\kappa\right)\left(\kappa g_0-g_1\dot{H}\tau
\right)}{\dot{H}\left(g_0+g_1\tau\right)^2\left(1+e^{-2H}\right)}dH\wedge dy\ ,
\end{align}
\eq
\tau=\frac{g_0}{g_1}\frac{\sqrt2(\dot H+\kappa)\left(1+e^{2H}\right)^{1/2}-
\sqrt{E_2}}{\sqrt2(\dot H+\kappa)\left(1+e^{2H}\right)^{1/2}+\sqrt{E_2}}\ .
\feq
Asymptotically for $H\to\infty$ the scalar field goes to its critical value,
$\tau\to g_0/g_1$, and the metric approaches AdS$_4$.
A more detailed analysis of the geometry \eqref{metr-expl} will be
presented elsewhere.

\subsection{$\texttt{G}_0=\psi_-=0$}
\label{G_0=psi_-=0}

For $\texttt{G}_0=\psi_-=0$ one has $\psi_1=\psi_1(z)$ by virtue of \eqref{thpsi07}
and \eqref{deriv-psi_1}. Moreover, the sum of \eqref{thpsi01} and \eqref{thpsi04}
yields
\eq
\psi_1=r\chi(w,\bar w)\ , \label{psi_1=rchi}
\feq
with $\chi(w,\bar w)$ an arbitrary function, while the difference of \eqref{thpsi01}
and \eqref{thpsi04} implies $A_z=\partial_z\varphi$. Subtracting \eqref{thpsi09} from
\eqref{thpsi08} leads to
\eq
\partial_z\ln r + 2i\left(\frac{\X}{\bar b}-\frac{\Xb}b\right) = 0\ . \label{dzlnr}
\feq
Plugging this into \eqref{thpsi02}, one gets $\partial_z\psi_2=0$.
Using equ.~\eqref{dzPhi} in \eqref{dzlnr}, we obtain $\partial_z\Phi=\partial_z\ln r$,
and thus
\eq
e^{\Phi} = r\Lambda(w,\bar w)\ , \label{PhiLambda}
\feq
where $\Lambda$ is again an arbitrary function. \eqref{thpsi12}, together with
$\partial_z\psi_2=0$, gives
\eq
\psi_2 = \frac{r^2}{\psi_1^2}\nu(w)\ , \label{psi_2}
\feq
with $\nu(w)$ holomorphic. Note that \eqref{psi_1=rchi}, combined with $\psi_1=\psi_1(z)$,
forces the phase $\theta$ of $\psi_1$ to be constant. By rescaling all the $\psi_{\texttt{i}}$'s
with $e^{-i\theta}$ we can thus choose $\psi_1$ real without loss of generality.
From the gaugino equations \eqref{htgIII1}-\eqref{htgIII4} one has
\eq
\partial_z z^{\alpha}=0\ , \qquad \psi_2\partial z^{\alpha} + \bar\psi_2\bar\partial
z^{\alpha} = 0\ , \label{dxz}
\feq
and hence $z^{\alpha}=z^{\alpha}(w,\bar w)$, $A_z=0=\partial_z\varphi$. In order to
proceed, it is convenient to distinguish two subcases, namely
$\X e^{i\varphi}-\Xb e^{-i\varphi}=0$ and $\X e^{i\varphi}-\Xb e^{-i\varphi}\neq 0$.

\subsubsection{$\X e^{i\varphi}-\Xb e^{-i\varphi}=0$}
\label{X-Xb=0}

If $\X e^{i\varphi}-\Xb e^{-i\varphi}=0$, \eqref{dzlnr} implies $r=r(w,\bar w)$.
Plugging this into \eqref{psi_1=rchi} and taking into account that $\psi_1=\psi_1(z)$,
we find that $\psi_1$ must be constant. By rescaling the $\psi_{\texttt{i}}$'s one can
then choose $\psi_1=1$ without loss of generality. Notice that \eqref{PhiLambda}
gives $\partial_z\Phi=0$ in this case, which is compatible with \eqref{dzPhi}.
From the sum of eqns.~\eqref{thpsi03} and \eqref{thpsi05} we get
\eq
A_w = \partial\varphi\ , \qquad A_{\bar w} = \bar\partial\varphi\ , \label{Avarphi}
\feq
whereas their difference leads to
\eq
\psi_2^{-1}e^{-2\Phi}\bar\partial\ln r = i\left(\frac{\X}{\bar b}+\frac{\Xb}b\right)\ .
\label{dbarlnr}
\feq
Taking the sum of \eqref{dbarlnr} and its complex conjugate, and using \eqref{psi_2},
one obtains
\eq
(\bar\nu(\bar w)\bar\partial+\nu(w)\partial)r = 0\ . \label{dxr}
\feq
Let us first consider the subcase $\psi_2\neq 0$, i.e., $\nu(w)\neq 0$. (The
case $\psi_2=0$ will be dealt with in section \ref{2=12=0}.) This allows to
introduce new coordinates $W,\bar W$ such that $\nu\partial=\partial_W$,
$\bar\nu\bar\partial=\partial_{\bar W}$. Using the residual gauge invariance
$w\mapsto W(w)$, $\Phi\mapsto\Phi-\frac12\ln(dW/dw)-\frac12\ln(d\bar W/d\bar w)$
leaving invariant the metric $e^{2\Phi}dwd\bar w$, one can set $\nu(w)=1$ and hence $w=W$
without loss of generality. Then, eqns.~\eqref{dxz} and \eqref{dxr} boil down to
\eq
\partial_z z^{\alpha} = \partial_x z^{\alpha} = \partial_x r = 0\ ,
\feq
where $x$ is defined by $w=x+iy$. Thus, $r=r(y)$, $z^{\alpha}=z^{\alpha}(y)$, $A_x=0$, and
from \eqref{Avarphi} also $\partial_x\varphi=0$ so that $\varphi=\varphi(y)$.
\eqref{dbarlnr} simplifies to
\eq
e^{-2\Phi}\partial_y r - 2r^2(\X e^{i\varphi}+\Xb e^{-i\varphi}) = 0\ . \label{dyr}
\feq
Plugging this into the sum of \eqref{thpsi08} and \eqref{thpsi09} yields
\eq
\partial e^{2\Phi} = \frac i{2r^5}\partial_y r\ , \label{partialPhi}
\feq
which implies $(\partial+\bar\partial)\Phi=0$, and thus $\Phi=\Phi(y)$. Integration
of \eqref{partialPhi} gives then
\eq
e^{2\Phi} = \frac1{4r^4}+L\ ,
\feq
with $L$ a real constant. In what follows, we shall use $r$ as a new coordinate in
place of $y$\footnote{This is possible as long as $\X\neq 0$, cf.~\eqref{dyr}.}.
The only nontrivial gaugino equation of the system \eqref{htgIII1}-\eqref{htgIII4}
becomes
\eq
r\frac{dz^{\alpha}}{dr} = \frac{g^{\alpha\bar\beta}{\cal D}_{\bar\beta}\Xb}{\Xb}\ .
\label{flow-psi_-=0}
\feq
One also has to check whether the equations \eqref{bianchi}-\eqref{Delta-Phi} for
the first Killing spinor are satisfied. The Bianchi identities \eqref{bianchi} and
Maxwell equations \eqref{maxwell} can be integrated once, with the result
\eq
\partial_y\left(\frac{X^I}{\bar b}-\frac{\bar X^I}b\right) = ip^I\ , \qquad
\partial_y\left(\frac{F_I}{\bar b}-\frac{\bar F_I}b\right) - \frac{ig_I}{r^4} = iq_I\ ,
\label{bianchi-max-psi_-=0}
\feq
where $p^I,q_I$ are integration constants. Using the flow equation \eqref{flow-psi_-=0}
together with the special geometry relation \cite{Vambroes}
\eq
-\frac12(\text{Im}\,{\cal N})^{-1|IJ} = \bar X^IX^J + g^{\alpha\bar\beta}{\cal D}_{\alpha}
X^I{\cal D}_{\bar\beta}\bar X^J\ , \label{ImN^-1}
\feq
one finds that \eqref{bianchi-max-psi_-=0}, as well as \eqref{Delta-Phi}, indeed hold,
if $p^I=0$, $q_I=4Lg_I$.

Finally, the shift vector $\sigma$ follows from \eqref{dsigma}, which implies
\begin{displaymath}
\sigma = \frac{dx}{4r^4}\ .
\end{displaymath}
Then the metric and the fluxes read respectively
\eq
ds^2 = -4r^2\left(dt+\frac{dx}{4r^4}\right)^2 + \frac{dz^2}{r^2} + \left(\frac1{4r^4}
+L\right)\frac{dx^2}{r^2} + \frac{dr^2}{16r^6\X\Xb\left(\frac1{4r^4}+L\right)}\ ,
\label{metr-psi_-=0}
\feq
\eq
F^I = -\frac2{\sqrt{\X\Xb}}(\text{Im}\,{\cal N})^{-1|IJ}g_J dt\wedge dr\ .
\label{fluxes-psi_-=0}
\feq
Actually the solutions with $L\neq 0$ can be cast into a simpler form by the
coordinate transformation
\begin{displaymath}
Lx = t-\psi\ , \qquad \zeta = |L|^{1/2}z\ , \qquad \rho^2 = \frac1{|L|r^2}\ .
\end{displaymath}
Defining also $q^2\equiv 4/|L|$, we get for $L>0$
\eq
ds^2 = -\left(\rho^2+\frac{q^2}{\rho^2}\right)dt^2 + \frac{d\rho^2}{4\X\Xb\left(\rho^2
+\frac{q^2}{\rho^2}\right)} + \rho^2(d\zeta^2+d\psi^2)\ , \label{naked}
\feq
and for $L<0$
\eq
ds^2 = \left(\rho^2-\frac{q^2}{\rho^2}\right)dt^2 + \frac{d\rho^2}{4\X\Xb\left(\rho^2
-\frac{q^2}{\rho^2}\right)} + \rho^2(d\zeta^2-d\psi^2)\ . \label{bubble}
\feq
In both cases, the fluxes and the flow equation \eqref{flow-psi_-=0} become
\eq
F^I = \frac q{\rho^2\sqrt{\X\Xb}}(\text{Im}\,{\cal N})^{-1|IJ}g_J dt\wedge d\rho\ ,
\qquad -\rho\frac{dz^{\alpha}}{d\rho} = \frac{g^{\alpha\bar\beta}{\cal D}_{\bar\beta}\Xb}
{\Xb}\ .
\feq
\eqref{naked} represents a generalization of the naked singularity solution to minimal
gauged supergravity found in \cite{Caldarelli:1998hg} with nontrivial scalars turned on.
Its double analytic continuation $t\mapsto it$, $\psi\mapsto i\psi$, $q\mapsto -iq$
yields \eqref{bubble}, which has the interpretation of a bubble of
nothing \cite{Witten:1981gj}: In order to avoid the conical singularity at
$\rho^2=q\equiv\rho_{\text s}^2$ in the $(t,\rho)$-hypersurface, we must compactify $t$
such that\footnote{We assumed that $\lim_{\rho\to\rho_{\text s}}g_IX^I(\rho)
\equiv X_{\text s}\neq 0$.}
\begin{displaymath}
t \sim t + \frac{\pi}{2\rho_{\text s}|X_{\text s}|}\ .
\end{displaymath}
Note that the limit $L\to0$ is naively singular in the coordinates $t,\rho,\zeta,\psi$,
because the charge $q$ diverges, but it can be taken if we perform a Penrose
limit \cite{Penrose}: Start for instance from the $L>0$ solution and set
\begin{displaymath}
\psi-t = -\epsilon^2X^+\ , \qquad \psi+t = 2X^-\ , \qquad \rho = \frac1{\epsilon R}\ ,
\qquad \zeta = \epsilon Z\ , \qquad q = \frac 2{\epsilon}\ .
\end{displaymath}
Then, the limit $\epsilon\to 0$ leads to the regular solution
\begin{displaymath}
ds^2 = -4R^2 dX^{-2} - \frac2{R^2}dX^-dX^+ + \frac{dR^2}{4R^2\X\Xb} + \frac{dZ^2}{R^2}\ ,
\end{displaymath}
\begin{displaymath}
F^I = -\frac2{\sqrt{\X\Xb}}(\text{Im}\,{\cal N})^{-1|IJ}g_J dX^-\wedge dR\ ,
\end{displaymath}
which is nothing else than \eqref{metr-psi_-=0} and \eqref{fluxes-psi_-=0} for $L=0$.

Integration of the system \eqref{httda0}-\eqref{htwbda0} yields
\begin{displaymath}
\psi_0 = \hat\psi_0 - \frac1{2r^2}\ ,
\end{displaymath}
with $\hat\psi_0$ a complex constant. The second Killing spinor is thus
\eq
\epsilon^1 = \left(\hat\psi_0-\frac1{2r^2}\right)1 + re^{\Phi}e_{12}\ , \qquad
\epsilon^2 = e^{\Phi-i\varphi}1 - \left(\frac1{2b}+\bar b\bar{\hat\psi}_0\right)e_{12}\ .
\label{2ndKill-psi_-=0}
\feq
For $\hat\psi_0=0$, the norm squared of the associated Killing vector
$V_{\mu}=A(\epsilon^i,\gamma_{\mu}\epsilon_i)$ (with $A$ given in \eqref{Majorana}) reads
\eq
V^2 = -4r^2L^2\ ,
\feq
which vanishes for $L=0$, so that in this case the solution belongs to the null class
as well. To understand what happens for $L\neq 0$, we have to consider a general linear
combination of the two Killing spinors. As was explained earlier, the rescaling
$(\epsilon^1,\epsilon^2)\mapsto(C\epsilon^1,\bar C\epsilon^2)$, with $C\in\bC$ an
arbitrary constant, gives again a Killing spinor. If we apply this to \eqref{2ndKill-psi_-=0}
and choose $\hat\psi_0=1/C$ (in order to recover the first covariantly constant spinor for
$C\to 0$), the associated Killing vector has norm squared
\eq
V^2 = -4r^2\left[(1+L|C|^2)^2 + \frac{\text{Im}^2C}{r^4}\right]\ .
\feq
This is zero iff $\text{Im}C=0$, $L=-1/|C|^2$, i.e. $L<0$. In conclusion, the half-BPS
solutions of this subsection belong also to the lightlike class for $L\le 0$. They must
therefore correspond to some of the geometries of \cite{Klemm:2009uw}, where the
half-supersymmetric null case was classified. This is indeed the case: Take the 1/2-BPS
solutions with $d\chi=0$ in section 5.2 of \cite{Klemm:2009uw}. Consider there the subcase
$d=\bar b\X/\Xb$, equ.~(5.49). In order to solve the equations for half-supersymmetry, make
the additional assumption that the function $H$, the scalars $z^{\alpha}$ and the wave profile
$\cal G$ depend on $w-\bar w$ only. Moreover, choose $m_J=g_J$ and $l^J=0$ in the expression
(5.67) that determines the fluxes. As a solution of the eqns.~(5.59), (5.62) for the wave
profile take ${\cal G}=-1/(4\rho^4)$. Finally, set $u=-2\sqrt2 t$, $v=-x/8$, $w+\bar w=\sqrt2 z$
and $\rho=1/r$. This yields the solution \eqref{flow-psi_-=0}, \eqref{metr-psi_-=0},
\eqref{fluxes-psi_-=0} with $L=0$. Note that for constant scalars, the $L=0$ solution reduces
to a subclass of the charged generalization of the Kaigorodov spacetime found in
\cite{Cai:2003xv}.

If one starts instead from the half-BPS null case with $d\chi\neq 0$, eqns.~(5.24), (5.33),
(5.34) in \cite{Klemm:2009uw}, and sets
\begin{displaymath}
u = A(t-Lx) + \frac z{\sqrt2 A}\ , \qquad v = A(t-Lx) - \frac z{\sqrt2 A}\ ,
\end{displaymath}
\begin{displaymath}
\Psi = -2^{7/4}At\ , \qquad \tanh\chi = \frac{\sqrt2 r^2}{A^2}\ ,
\end{displaymath}
where $A=(2|L|)^{-1/4}$, one obtains the $L<0$ solution. Notice that the geometry described
by eqns.~(5.24), (5.33) and (5.34) of \cite{Klemm:2009uw} appeared also in subsection
\ref{psi_-neq0}.

\subsubsection{$\X e^{i\varphi}-\Xb e^{-i\varphi}\neq 0$}
\label{X-Xbneq0}

For $\X e^{i\varphi}-\Xb e^{-i\varphi}\neq 0$, taking into account that the scalar fields
$z^{\alpha}$ and the phase $\varphi$ are independent of $z$, integration of \eqref{dzlnr}
yields
\eq
r = 2iz(\Xb e^{-i\varphi}-\X e^{i\varphi})\ , \label{r}
\feq
where a possible integration constant has been eliminated by shifting $z$. Using this
in \eqref{psi_1=rchi} and keeping in mind that $\psi_1$ depends on $z$ only, one gets
$\psi_1=cz$, with $c$ a real integration constant that we can set equal to one without
loss of generality by rescaling the $\psi_{\texttt{i}}$'s.
Plugging \eqref{r} into \eqref{PhiLambda}, we have $e^\Phi=ze^H$, with the real
function $H(w,\bar w)$ given by
\begin{displaymath}
e^H = 2i(\Xb e^{-i\varphi} - \X e^{i\varphi})\Lambda(w,\bar w)\ .
\end{displaymath}
From \eqref{psi_2} one obtains
\begin{displaymath}
\psi_2=-4\nu\left(\Xb e^{-i\varphi} - \X e^{i\varphi}\right)^2\ .
\end{displaymath}
In what follows, it is convenient to introduce the real function $Y=Y(w,\bar w)$,
\begin{equation}
Y=-i\frac{e^{i\varphi}\X+e^{-i\varphi}\Xb}{e^{i\varphi}\X-e^{-i\varphi}\Xb}\ ,
\end{equation}
which is related to the phase $\varphi$ of $b$ by
\[
e^{2i\varphi}=-\frac{1+iY}{1-iY}\frac{\Xb}{\X}\ .
\]
In terms of $Y$, the expressions for $\psi_2$ and $b$ simplify to
\begin{equation}
\psi_2=\frac{16\X\Xb}{1+Y^2}\nu\ , \qquad b=\frac{4i\Xb}{1-iY}z\ .
\end{equation}
The system \eqref{thpsi01}-\eqref{thpsi12} boils down to
\begin{eqnarray}
\label{gravI}e^{2H}\nu&=&-\frac{i}{8\X\Xb}\left[\bar\partial
Y-\frac{1+Y^2}{2Y}\bar\partial\ln\left(\X\Xb\right)\right]\ ,\\
\label{gravII}\partial\left(e^{2H}\nu\right)&=&-\frac{ie^{2H}Y(1+Y^2)}{32\X\Xb}\ ,
\end{eqnarray}
together with
\begin{displaymath}
A_w=\frac1{2Y}\left[\left(1+iY\right)\partial\ln
(\X)+\left(1-iY\right)\partial\ln (\Xb)\right]\ .
\end{displaymath}
Equ.~(\ref{Delta-Phi}) becomes
\begin{equation}
2\partial\bar\partial H=e^{2H}\left[\frac
12+Y^2+\frac{1+Y^2}{8\X\Xb}\left(\mbox{Im}\,\mathcal{N}\right)^{-1|IJ}g_Ig_J\right]\ .
\label{Delta-H}
\end{equation}
Using
\[
\left(\mbox{Im}\,\mathcal{N}\right)^{-1|IJ}g_Ig_J=-2\X\Xb+\frac{i\left(1+Y^2\right)}
{8e^{2H}Y\bar\nu}\partial\ln (\X\Xb)\ ,
\]
that follows from \eqref{ImN^-1}, it is easy to shew that \eqref{Delta-H} is
automatically satisfied if \eqref{gravI} and \eqref{gravII} hold.

The case $\nu=0$ (and thus $\psi_2=\psi_{12}=0$) will be considered in \ref{2=12=0}.
In the remaining part of this subsection we shall assume $\nu\neq0$, which allows to
define new coordinates $W$, $\bar W$ such that
\begin{displaymath}
\partial_W=\nu\partial\ , \qquad \partial_{\bar W}=\bar\nu\bar\partial\ .
\end{displaymath}
Making use of the residual gauge invariance $w\mapsto W(w)$,
$\Phi\mapsto\Phi-\frac12\ln(dW/dw)-\frac12\ln(d\bar W/d\bar w)$ leaving invariant
the metric $e^{2\Phi}dwd\bar w$, one can set $\nu(w)=1$ and hence $w=W$ without loss of
generality. The gaugino eqns.~\eqref{htgIII1} and \eqref{htgIII4} reduce to
\eq
(\partial+\bar\partial)z^\alpha = 0\ , \qquad
\partial z^\alpha = -\frac{8e^{2H}\X }{1+iY}g^{\alpha\bar\beta}\mathcal{D}_{\bar\beta}\Xb\ ,
\label{gaug}
\feq
which imply that $z^\alpha=z^\alpha(w-\bar w)$. Note also that from \eqref{gravII}
it follows that the functions $H$, $Y$ depend on $w-\bar w$ only.

The Bianchi identities (\ref{bianchi}) and Maxwell equations
\eqref{maxwell} are automatically satisfied. Finally, integration of \eqref{dsigma}
gives the shift vector
\begin{equation}
\sigma=\frac{e^{2H}}{2z}(dw+d\bar w )\ .
\end{equation}
Denoting with a dot the derivative w.r.t.~$i(w-\bar w)$, \eqref{gaug}, \eqref{gravI}
and \eqref{gravII} become
\begin{eqnarray}
\label{scal1}\dot z^\alpha&=&\frac{8ie^{2H}\X }{1+iY}
g^{\alpha\bar\beta}\mathcal{D}_{\bar\beta}\Xb\ , \\
\label{Y1}e^{2H}&=&-\frac1{8\X\Xb}\left\{\dot Y-\frac{1+Y^2}{2Y}
\left[\ln\left(\X\Xb\right)\right]^{\!\text{\large{$\cdot$}}}\right\}\ , \\
\label{Y2}\dot H&=&-\frac{Y(1+Y^2)}{64\X\Xb}\ .
\end{eqnarray}
Combining (\ref{Y1}) and (\ref{Y2}) yields
\begin{eqnarray}
\left[\frac{\dot Y}{\X\Xb}\right]^{\!\text{\large{$\cdot$}}}&=&-\frac{Y\left(1+Y^2\right)}
{32\left(\X\Xb\right)^2}\left\{\dot Y-\frac{1+Y^2}{2Y}\left[\ln\left(\X\Xb\right)
\right]^{\!\text{\large{$\cdot$}}}\right\}\nonumber\\
&&+\left\{\frac{1+Y^2}{2Y}\frac{\left[\ln\left(\X\Xb\right)\right]^{\!\text{\large{$\cdot$}}}}
{\X\Xb}\right\}^{\!\text{\large{$\cdot$}}}\ ,
\end{eqnarray}
which, integrated once, gives
\begin{equation}
\left(\ln\frac{\X\Xb}{1+Y^2}\right)^{\!\text{\large{$\cdot$}}} = \frac{Y(1+Y^2)}{64\X\Xb}
- \frac{64YL\X\Xb}{1+Y^2}\ ,
\end{equation}
where $L$ is a real integration constant. Let us define
\begin{displaymath}
e^{\xi} = \frac{64\X\Xb}{1+Y^2}\ ,
\end{displaymath}
and use $\xi$ as a new coordinate instead of $w-\bar w$. Then, the flow equation
\eqref{scal1} becomes
\eq
\frac{dz^{\alpha}}{d\xi} = \frac i{2\Xb Y}(1-iY)g^{\alpha\bar\beta}{\cal D}_{\bar\beta}\Xb\ ,
\label{dzdxi}
\feq
with $Y$ given by $Y^2=64e^{-\xi}\X\Xb-1$. Setting $x=(w+\bar w)/2$, the metric and the
fluxes read respectively
\begin{eqnarray}
ds^2 &=& -z^2e^{\xi}\left[dt+4(e^{-2\xi}-L)\frac{dx}z\right]^2 + 4e^{-\xi}\frac{dz^2}{z^2}
\nonumber \\
&& \qquad +16e^{-\xi}(e^{-2\xi}-L)dx^2 + \frac{4e^{-2\xi}d\xi^2}{Y^2(e^{-\xi}-Le^{\xi})}\ ,
\label{metr-Y}
\end{eqnarray}
\begin{eqnarray}
F^I&=&8i\left(\frac{\Xb X^I}{1-iY}-\frac{\X \bar X^I}{1+iY}\right)dt\wedge dz \\
&&+\frac 4Y\left[\frac{2\Xb X^I}{1-iY}+\frac{2\X\bar X^I}{1+iY}+\left(\mbox{Im}\,\mathcal{N}
\right)^{-1|IJ}g_J\right](zdt-4Ldx)\wedge d\xi\ . \nonumber
\end{eqnarray}
For $L>0$, the line element \eqref{metr-Y} can be cast into the simple form
\begin{eqnarray}
ds^2 &=& 4e^{-\xi}\left(-z^2d{\hat t}^2 + \frac{dz^2}{z^2}\right) + 16L(e^{-\xi}-Le^{\xi})
\left(dx - \frac z{2\sqrt L}d\hat t\right)^2 \nonumber \\
&& \qquad + \frac{4e^{-2\xi}d\xi^2}{Y^2(e^{-\xi}-Le^{\xi})}\ , \label{near-hor}
\end{eqnarray}
where $\hat t\equiv t/(2\sqrt L)$. \eqref{near-hor} is of the form (3.3) of
\cite{Astefanesei:2006dd}, and describes the near-horizon geometry of extremal
rotating black holes. From \eqref{dzdxi} it is clear that the scalar fields have
a nontrivial dependence on the horizon coordinate $\xi$ unless ${\cal D}_{\alpha}\X=0$.
While the generic hairy black holes with the near-horizon geometry \eqref{near-hor}
are still to be discovered, the solution with constant scalars is actually known:
Start from the rotating generalization of the hyperbolic black hole solution
to minimal gauged supergravity, given by \cite{Caldarelli:1998hg}
\begin{displaymath}
ds^2 = -\frac{\Delta_r}{\rho^2}\left[dt + \frac a{\Xi}\sinh^2\!\theta d\phi\right]^2
+ \frac{\rho^2}{\Delta_r}dr^2 + \frac{\rho^2}{\Delta_{\theta}}d\theta^2 +
\frac{\Delta_{\theta}\sinh^2\!\theta}{\rho^2}\left[adt - \frac{r^2+a^2}{\Xi}d\phi\right]^2\ ,
\end{displaymath}
\begin{displaymath}
A = -\frac{q_{\text e}r}{\rho^2}\left[dt + \frac a{\Xi}\sinh^2\!\theta d\phi\right] -
\frac{q_{\text m}\cosh\theta}{\rho^2}\left[adt - \frac{r^2+a^2}{\Xi}d\phi\right]\ ,
\end{displaymath}
with
\begin{displaymath}
\Delta_r = (r^2+a^2)\left(-1+\frac{r^2}{\ell^2}\right) - 2mr + q_{\text e}^2 +
q_{\text m}^2\ , \qquad \Delta_{\theta} = 1 + \frac{a^2}{\ell^2}\cosh^2\!\theta\ ,
\end{displaymath}
\begin{displaymath}
\rho^2 = r^2 + a^2\cosh^2\!\theta\ , \qquad \Xi = 1 + \frac{a^2}{\ell^2}\ .
\end{displaymath}
Here, $a$, $m$, $q_{\text e}$ and $q_{\text m}$ denote the rotation parameter, mass
parameter, electric and magnetic charge respectively, and $\ell$ is related to
the cosmological constant by $\Lambda=-3/\ell^2$. This black hole is both extremal and
supersymmetric iff \cite{Caldarelli:1998hg}
\eq
m = q_{\text e} = 0\ , \qquad q_{\text m} = \pm\frac{\ell}2\Xi\ ,
\feq
which leaves a one-parameter family of solutions, with
horizon at $r^2=r_{\text h}^2=(\ell^2-a^2)/2$. In order to obtain the
near-horizon limit, we introduce new coordinates $z$, $\hat t$, $\hat\phi$
according to
\eq
r = r_{\text h} + \epsilon r_0z\ , \qquad t = \frac{\hat t r_0}{\epsilon}\ , \qquad
\phi = \hat\phi + \Omega\frac{\hat t r_0}{\epsilon}\ ,
\feq
where $\Omega=a\Xi/(r_{\text h}^2+a^2)$ is the angular velocity of the horizon, and $r_0$
is defined by
\begin{displaymath}
r_0^2 = \frac{\ell^2(r_{\text h}^2+a^2)}{4r_{\text h}^2}\ .
\end{displaymath}
After taking the limit $\epsilon\to 0$, the metric becomes
\eq
ds^2 = \frac{\ell^2\rho_{\text h}^2}{4r_{\text h}^2}\left[-z^2d{\hat t}^2 +
\frac{dz^2}{z^2}\right] + \frac{\rho_{\text h}^2}{\Delta_{\theta}}d\theta^2
+ \frac{\Delta_{\theta}\sinh^2\!\theta}{\rho_{\text h}^2\,\Xi^2}(r_{\text h}^2+a^2)^2(d\hat\phi
+ kzd\hat t)^2\ , \label{near-hor-hyp}
\feq
with
\begin{displaymath}
\rho_{\text h}^2 = r_{\text h}^2 + a^2\cosh^2\!\theta\ , \qquad
k = \frac{2r_{\text h}r_0^2\Omega}{r_{\text h}^2+a^2}\ .
\end{displaymath}
If we set
\begin{displaymath}
e^{-\xi} = \frac{\ell^2\rho_{\text h}^2}{16r_{\text h}^2}\ , \qquad x =
-\frac{32r_{\text h}^3(r_{\text h}^2+a^2)}{\ell^6\,\Xi^2a}\hat\phi\ , \qquad
L = \frac{\ell^8\Xi^2}{1024r_{\text h}^4}\ , \qquad \X\Xb = \frac1{4\ell^2}\ ,
\end{displaymath}
\eqref{near-hor} reduces precisely to the near-horizon geometry \eqref{near-hor-hyp}.

Let us now come back to the case of arbitrary $L$.
The missing component $\psi_0$ of the second Killing spinor is determined by the
system \eqref{httda0}-\eqref{htwbda0}, that simplifies to
\eq
\partial_t\psi_0=1\ , \qquad \partial_z\psi_0=\frac{1+Y^2}{32z^2\X\Xb}\ , \qquad
\partial\psi_0=-\bar\partial\psi_0=\frac{ie^{2H}Y}{2z}\ . \label{psi_0Y}
\feq
Integration of \eqref{psi_0Y} yields (after going back to the original basis)
\begin{align}
\alpha&=\hat\alpha+t-\frac{1+Y^2}{32z\X\Xb}\ , \qquad
\beta=-\frac{4i\X e^H}{1+iY}e^{i\varphi}\ , \nonumber\\
\gamma&=\frac{e^H}{z}e^{-i\varphi}\ , \qquad
\delta=\frac{4i\X}{1+iY}z\left(\bar{\hat\alpha}+t\right)-\frac{1-iY}{8i\Xb}\ ,
\end{align}
where $\hat\alpha\in\bC$ denotes an integration constant. As before, we rescale
$\alpha,\beta$ by $C$ and $\gamma,\delta$ by $\bar C$, with $C\in\bC$ constant,
and choose $\hat\alpha=1/C$ in order to obtain the first Killing spinor for $C\to 0$.
Then, the norm squared of the associated Killing vector turns out to be
\begin{align}
V^2=&-4|b|^2\left[|1+C
t|^2+\left(\frac{e^{2H}}{z^2}-\frac{z^2}{4|b|^4}\right)|C|^2\right]^2
-\left(\frac{2z\mbox{Im}C}{|b|}\right)^2\ ,
\end{align}
which is always negative, so that the solutions considered here do not belong
to the null class\footnote{Of course, the choice $\hat\alpha=1/C$ does not cover
the case $\hat\alpha=0$, which has to be treated separately. It is easy to show
that the result is again a timelike vector.}.

Notice that in minimal supergravity, the analogue of eqns.~\eqref{Y1},
\eqref{Y2} follow from the dimensionally reduced gravitational Chern-Simons
action \cite{Cacciatori:2007vn}. It would be interesting to see if something similar
happens here. For instance, \eqref{scal1}-\eqref{Y2} might be related to the
gravitational Chern-Simons system coupled to scalar fields. We hope to come
back to these points in a future publication.

\subsubsection{$\psi_2=0$}\label{2=12=0}

In \ref{X-Xb=0} and \ref{X-Xbneq0} we assumed $\nu\neq0$, that is $\psi_2\neq0$.
Let us now consider the case $\texttt{G}_0=0$ and $\psi_2=\psi_{12}=0$. The gaugino
equations \eqref{htgIII1}-\eqref{htgIII4} imply that the scalars $z^\alpha$ are constant, while the
system \eqref{httda0}-\eqref{htwbda0} and \eqref{thpsi01}-\eqref{thpsi12} reduces to
\begin{eqnarray}\label{2=12=0psi0}
\partial_t\psi_0&=&-\frac{4i\X}{\bar b}\psi_1\ , \qquad
\partial_z\psi_0=\frac{\partial_t\psi_0}{2r^2}\ , \qquad
\partial\psi_0=\sigma_w\partial_t\psi_0\ , \\
\bar\partial\psi_0&=&\sigma_{\bar w}\partial_t\psi_0\ , \qquad
\psi_1=\psi_1(z)\ , \qquad \partial_z\psi_1=\frac{4i\Xb}{b}\psi_1\ ,
\end{eqnarray}
together with
\begin{equation}\label{betaeq}
\partial_zr=-4i\X e^{i\varphi}\ , \qquad
\partial r=\partial\varphi=\partial_z\varphi=0\ , \qquad
e^{i\varphi}\X+e^{-i\varphi}\Xb=0\ .
\end{equation}
From \eqref{dsigma} one gets $\sigma=0$, and (\ref{2=12=0psi0})-(\ref{betaeq}) give
\begin{equation}
\psi_0=\hat{\alpha}+t-\frac1{32z\X\Xb}\ , \qquad \psi_1=z\ , \qquad
b=4i\Xb z\ ,
\end{equation}
where $\hat{\alpha}\in\bC$ is an integration constant. It is straightforward to shew
that the Killing vector associated to a general linear combination of the two
Killing spinors is always timelike. Integration of (\ref{dzPhi}) yields $e^\Phi=ze^H$,
with $H=H(w,\bar w)$ a real function satisfying
\begin{equation}
8\partial\bar\partial H=e^{2H} \label{Liouville-H}
\end{equation}
due to \eqref{Delta-Phi}. \eqref{Liouville-H} is the Liouville equation and implies
that the two-dimensional metric $e^{2H}dwd\bar w$ has constant negative curvature.
Note that the Bianchi identities \eqref{bianchi} and Maxwell equations
\eqref{maxwell} are automatically satisfied.
The metric and  fluxes read respectively
\begin{eqnarray}
ds^2&=&-64\X\Xb z^2dt^2+\frac{dz^2}{16\X\Xb z^2}+\frac{e^{2H}dwd\bar
w}{16\X\Xb}\ ,\\
F^I&=&-16{\text{Im}}(\Xb X^I)dt\wedge dz + \frac{ie^{2H}}{16\X\Xb}\left[4{\text{Re}}
(\Xb X^I)+g_J\left({\text{Im}}\,{\cal N}\right)^{-1|IJ}\right]dw\wedge d\bar w\ .
\nonumber
\end{eqnarray}
We have thus a product spacetime AdS$_2$ $\times$ H$^2$, with constant electric
flux on AdS$_2$ and magnetic flux on H$^2$. This is the near-horizon geometry of
static supersymmetric black holes, like the ones discovered in \cite{Cacciatori:2009iz}.

\subsection{Case $\texttt{G}_0\neq 0$}

For $\texttt{G}_0\neq 0$, the gaugino eqns.~\eqref{htgIII1}-\eqref{htgIII4}
suggest to define new coordinates $Z,W,\bar W$ according to
\eq
z=z(Z,W,\bar W)\ , \qquad w=W\ , \qquad \bar w=\bar W\ ,
\feq
where
\eq
\frac{\partial z}{\partial W} = -\frac{\psi_1}{2\psi_-}\ . \label{dz/dW}
\feq
Then, \eqref{htgIII2} and \eqref{htgIII3} simplify to
\eq
\partial_{\bar W}z^{\alpha} = \partial_Wz^{\alpha} = 0\ ,
\feq
so that the scalars depend on $Z$ only. The integrability conditions
\begin{displaymath}
\frac{\partial^2z}{\partial\bar W\partial W} =
\frac{\partial^2z}{\partial W\partial\bar W}\ ,
\end{displaymath}
of \eqref{dz/dW} and its complex conjugate read
\eq
\partial_{\bar W}\frac{\psi_1}{\psi_-} = \partial_W\frac{\bar\psi_1}
{\bar\psi_-}\ . \label{int-cond-W}
\feq
Remarkably, it can be shown that \eqref{int-cond-W} is implied by the
Killing spinor eqns.~\eqref{thpsi01}-\eqref{thpsi12}. Unfortunately,
the system \eqref{thpsi01}-\eqref{thpsi12} does not seem to simplify
much after the introduction of the coordinates $Z,W,\bar W$, at least not
in an obvious way, so that we were unable to solve it in general in the
case $\texttt{G}_0\neq 0$. For minimal ${\cal N}=2$ gauged supergravity,
all known 1/2 BPS solutions have either $\texttt{G}_0=0$, or are related
to the case $\texttt{G}_0=0$ by a diffeomorphism \cite{Cacciatori:2007vn}.
This might be a general feature, and hold in the matter-coupled case as
well, but we know of no way to show this in general.

\acknowledgments

This work was partially supported by INFN and MIUR-PRIN contract 20075ATT78.

\normalsize

\appendix

\section{Spinors and forms}
\label{spinors}

In this appendix, we summarize the essential information needed to realize
spinors of Spin(3,1) in terms of forms (cf.~also \cite{Gillard:2004xq}
and references therein).

Let $V = \bR^{3,1}$ be a real vector space equipped with the Lorentzian inner
product $\langle\cdot,\cdot\rangle$. Introduce an orthonormal basis $e_1, e_2, e_3, e_0$,
where $e_0$ is along the time direction, and consider the subspace $U$
spanned by the first two basis vectors $e_1, e_2$. The space of Dirac spinors
is $\Delta_c = \Lambda^{\ast}(U\otimes \bC)$, with basis
$1, e_1, e_2, e_{12} = e_1 \wedge e_2$.
The gamma matrices are represented on $\Delta_c$ as
\eqn
\gamma_{0}\eta&=&-e_2\wedge\eta+e_2\rfloor\eta\,, \qquad
\gamma_{1}\eta=e_1\wedge\eta+e_1\rfloor\eta\,, \nonumber \\
\gamma_{2}\eta&=&e_2\wedge\eta+e_2\rfloor\eta\,, \qquad
\gamma_{3}\eta=ie_1\wedge\eta-ie_1\rfloor\eta\,,
\feqn
where
\begin{displaymath}
\eta = \frac 1{k!}\eta_{j_1\ldots j_k} e_{j_1}\wedge\ldots\wedge e_{j_k}
\end{displaymath}
is a $k$-form and
\begin{displaymath}
e_i\rfloor\eta = \frac 1{(k-1)!}\eta_{ij_1\ldots j_{k-1}} e_{j_1}\wedge\ldots
                  \wedge e_{j_{k-1}}\,.
\end{displaymath}
One easily checks that this representation of the gamma matrices satisfies
the Clifford algebra relations $\{\gamma_a, \gamma_b\} = 2\eta_{ab}$.
The parity matrix is defined by $\gamma_5 = i\gamma_0\gamma_1\gamma_2\gamma_3$,
and one finds that the even forms $1, e_{12}$ have positive chirality,
$\gamma_5\eta = \eta$, while the odd forms $e_1, e_2$ have negative chirality,
$\gamma_5\eta = -\eta$, so that $\Delta_c$ decomposes into two complex chiral
Weyl representations $\Delta_c^+ = \Lambda^{\mathrm{even}}(U\otimes \bC)$ and
$\Delta_c^- = \Lambda^{\mathrm{odd}}(U\otimes \bC)$. \\
Let us define the auxiliary inner product
\begin{equation}
\langle\sum_{i=1}^2 \alpha_i e_i, \sum_{j=1}^2 \beta_j e_j\rangle = \sum_{i=1}^2
\alpha_i^{\ast}\beta_i
\end{equation}
on $U\otimes \bC$, and then extend it to $\Delta_c$. The Spin(3,1) invariant
Dirac inner product is then given by
\begin{equation}
D(\eta, \theta) = \langle\gamma_0\eta, \theta\rangle\,.
\end{equation}
The Majorana inner product that we use is
\begin{equation}
A(\eta, \theta) = \langle C\eta^{\ast}, \theta\rangle\,, \label{Majorana}
\end{equation}
with the charge conjugation matrix $C=\gamma_{12}$. It is easy to show \cite{Cacciatori:2008ek}
that \eqref{Majorana} is Spin(3,1) invariant as well.

A Killing spinor can be viewed as an SU(2) doublet
$(\epsilon^1, \epsilon^2)$, where an upper index means that a spinor has positive
chirality. $\epsilon^i$ is related to the negative chirality spinor $\epsilon_i$ by
charge conjugation, $\epsilon_i^C = \epsilon^i$, with
\begin{equation}
\epsilon_i^C = \gamma_0 C^{-1}\epsilon_i^{\ast}\,.
\end{equation}
As was shown in \cite{Cacciatori:2008ek}, there are three orbits of spinors under
Spin(3,1), two of them with corresponding null bilinear
$V_{\mu}=A(\epsilon^i,\gamma_{\mu}\epsilon_i)$, and one with timelike $V_{\mu}$.
In the latter case, one can choose $(\epsilon^1,\epsilon_2)=(1,be_2)$ as
representative \cite{Cacciatori:2008ek}, with $b$ a complex-valued function.

\section{The case $D_zP+ie^{-2\Phi}b^2F_{zw}=0$}
\label{DzP}

In section \ref{time-dep-Kill}, we simplified the equations for the second
Killing spinor under the assumption $D_zP+ie^{-2\Phi}b^2F_{zw}\neq0$. Here
we consider the case $D_zP+ie^{-2\Phi}b^2F_{zw}=0$. From \eqref{httw2}, one
obtains then $DP=0$ or $\psi_-=0$. Let us first assume the latter, i.e.,
$\psi_2=\psi_{12}$. Then, the $\partial_t\beta$, $\partial_t\bar\gamma$,
$\partial_z\beta$, $\partial_z\bar\gamma$, $\partial\beta$ and
$\partial\bar\gamma$ eqns.~of \eqref{htgr1}-\eqref{htgr4} imply
\begin{equation}\label{htgpsi-=0c3}
\partial_z\psi_2=2\left[\partial_z\ln r+2i\left(\frac{\X}{\bar b}-\frac{\Xb}b
\right)\right]\psi_2\ , \qquad \partial_t\psi_2=0\ ,
\end{equation}
\begin{equation}\label{htgpsi-=0c2}
\left[\partial_z\ln r+2i\left(\frac{\X}{\bar b}-\frac{\Xb}b\right)\right]
\psi_1=0\ ,
\end{equation}
\begin{eqnarray}
e^{-2\Phi}(\bar\partial\ln r)\psi_1-i\left(\frac{\X}{\bar b}+\frac{\Xb}b\right)
\psi_2&=&0\ , \label{htgpsi-=0c1} \\
e^{-2\Phi}\left(A_{\bar w}-\bar\partial\varphi\right)\psi_1+\left(\frac{\X}
{\bar b}-\frac{\Xb}b\right)\psi_2&=&0\ . \label{htgpsi-=0c0}
\end{eqnarray}
We have to suppose $\psi_1\neq0$ because otherwise (\ref{htgpsi-=0c1}) and
\eqref{htgpsi-=0c0} lead to $\psi_2=0$\footnote{This is true if $\X\neq 0$.}
and thus there exists no further Killing spinor. Hence, \eqref{htgpsi-=0c3}
and \eqref{htgpsi-=0c2} yield $\partial_z\psi_2=0$. Deriving
\eqref{htgpsi-=0c1} and \eqref{htgpsi-=0c0} with respect to $t$ we get
\[
0=\bar\partial r\partial_t\psi_1\ , \qquad 0=\left(A_{\bar w}-\bar\partial
\varphi\right)\partial_t\psi_1\ .
\]
If $\partial_t\psi_1\neq0$ then $\bar\partial r=0$,
$\bar\partial\varphi=A_{\bar w}$ and \eqref{htgpsi-=0c1}, \eqref{htgpsi-=0c0}
give $\psi_2=0$. The gaugini equations (\ref{htgIII1})-(\ref{htgIII4}) imply
then that the scalar fields $z^\alpha$ must be constant. Moreover, since
in this case $A_{\mu}=0$, one has also $\partial\varphi=\bar\partial\varphi=0$,
which, together with $\partial r=\bar\partial r=0$ leads to $b=b(z)$.

If instead $\partial_t\psi_1=0$, all the $\psi_{\texttt{i}}$,
$\texttt{i}=1,2,12$, are independent of $t$, and the Killing spinor equations
reduce to the system \eqref{thpsi01}-\eqref{thpsi12} with $\texttt{G}_0=0$
and $\psi_-=0$, which is solved in section \ref{G_0=psi_-=0}.\\

In the case $DP=0$, consider the integrability condition (\ref{httw1}). As
long as $D_zQ-ie^{-2\Phi}\bar b^2F_{zw}\neq0$ one could proceed exactly in the
same way as in section \ref{time-dep-Kill}. If
$D_zQ-ie^{-2\Phi}\bar b^2F_{zw}=0$, (\ref{httw1}) implies $\psi_-=0$ or $DQ=0$.
The case $\psi_-=0$ was already considered above, so the only remaining case is
\[
D_zP+ie^{-2\Phi}b^2F_{zw}=DP=D_zQ-ie^{-2\Phi}\bar b^2F_{zw}=DQ=0\ .
\]
For minimal gauged supergravity, one can show \cite{Cacciatori:2007vn} that
this brings us back again to the case $\psi_-=0$. Perhaps an analogous
reasoning can be applied here as well, although we shall not attempt to do this.

\end{document}